%% file: AdaptiveFarmerJoshi.tex
\documentclass[final,5p,twocolumn]{elsarticle}
\include{Preamble}

\begin{document}
\begin{filecontents*}{"Data/AFJ-Moments.txt"}
Moment,GA,Lower_GA,Upper_GA,NMTA,Lower_NMTA,Upper_NMTA,Emperical
mean,-0.00003,-0.00082,0.00055,-0.00011,-0.00096,0.00054,-0.00005
stdev,0.02594,0.02102,0.02741,0.02389,0.02035,0.02589,0.02767
kurtosis,6.79172,0.46553,5.62495,2.67784,0.45883,2.88263,4.03912
ks,0.01675,0.01637,0.04111,0.02322,0.01637,0.04492,0
hurst,0.54975,0.47187,0.56053,0.55758,0.48541,0.57422,0.54487
gph,0.35897,0.14932,0.60416,0.43850,0.10793,0.60692,0.73580
adf,-55.72727,-55.40820,-48.46334,-51.56574,-55.05093,-47.79636,-49.42858
garch,0.95984,0.94076,0.97653,0.95906,0.92315,0.97948,0.99339
hill,0.39393,0.26899,0.38734,0.34828,0.26658,0.37113,0.42845
\end{filecontents*}

\begin{filecontents*}{"Data/AFJ-Parameters.txt"}
Parameter,GA,Lower_GA,Upper_GA,NMTA,Lower_NMTA,Upper_NMTA
N,90,80,121,81,101,182
lambda,10.27527,10.56002,12.62532,3.26799,5.47222,12.08974
a,0.08186,0.07760,0.11010,0.01525,0.01849,0.08230
d_max,87,88,113,105,86,129
d_min,34,37,53,90,30,67
mu_eta,-0.00255,-0.00211,0.00299,-0.00596,-0.00402,0.00282
sigma_eta,0.02767,0.01712,0.03046,0.04145,0.01212,0.03439
sigma_zeta,0.01721,0.01769,0.02103,0.01897,0.01399,0.01775
T_max,0.67289,0.60669,0.98115,1.414,0.82567,1.49038
tau_min,-0.39869,-0.52035,-0.39311,-0.52416,-0.4843,-0.29535
v_max,0.43789,0.06897,0.34144,-0.07734,0.09896,0.43791
v_min,-0.094,-0.31161,-0.16015,-0.38161,-0.25692,-0.06936
Gamma,0.57012,0.38411,0.59631,0.12921,0.18155,0.53080
H,45,43,57,54,26,72
T_min,0.35379,0.19958,0.47418,0.60891,0.28719,0.74903
tau_max,0.15383,0.0979,0.32634,-0.13256,-0.00543,0.12641
Fitness,-20.64028,-65.01577,-19.19762,-16.06043,-25.41958,-16.87596
\end{filecontents*}

\begin{filecontents*}{"Data/SFJ-Moments.txt"}
Moment,GA,Lower_GA,Upper_GA,NMTA,Lower_NMTA,Upper_NMTA,Emperical
mean,0.00025,-0.00040,0.00150,0.00112,-0.00021,0.00133,-0.00005
stdev,0.02052,0.0.02096,0.14332,0.02054,0.01947,0.13265,0.02767
kurtosis,0.26094,0.08238,14.06729,1.07280,0.00489,5.33540,4.03912
ks,0.03997,0.02589,0.18995,0.04263,0.03388,0.16749,0
hurst,0.58085,0.42944,0.59136,0.58833,0.44770,0.60538,0.54487
garch,0.98393,0.02273,1.31436,0.33447,0.00555,1.26677,0.99339
gph,0.25637,0.07374,0.72297,0.19139,-0.03422,0.62494,0.73580
adf,-0.53381,-53.73332,-13.33092,-52.55660,-53.53643,-16.19956,-49.42858
hill,0.26949,0.24877,0.77339,0.29442,0.24374,0.53700,0.42845
\end{filecontents*}

\begin{filecontents*}{"Data/SFJ-Parameters.txt"}
Parameter,GA,Lower_GA,Upper_GA,NMTA,Lower_NMTA,Upper_NMTA
N,84,80,120,61,48,203
lambda,12.59549,9.06162,11.44419,10.26531,7.42702,13.40640
a,0.45737,0.25131,0.37690,0.55884,0.16518,0.68659
d_max,49,57,80,64,44,92
d_min,21,26,46,18,11,39
mu_eta,0.00853,-0.00501,0.00361,0.00561,-0.00496,0.00319
sigma_eta,0.04200,0.02619,0.03708,0.02176,0.02705,0.05711
sigma_zeta,0.02032,0.01896,0.02009,0.01930,0.01738,0.02031
T_max,0.96590,1.00311,1.19918,1.47698,1.27609,1.78699
tau_min,-0.16188,-0.3251,-0.19622,-0.13981,-0.51354,-0.15336
v_max,0.25206,-0.01738,0.31747,0.13012,-0.02491,0.52813
v_min,-0.27687,-0.32853,-0.24581,-0.21514,-0.32808,-0.08448
T_min,0.40217,0.50307,0.67188,0.50747,0.46721,0.64138
tau_max,0.63070,0.3515,0.55426,0.60403,0.15530,0.76696
Fitness,-30.54963,-37.56189,-33.08459,-22.64955,-42.19412,-26.26681
\end{filecontents*}

\begin{frontmatter}
    \title{Calibrating an adaptive Farmer-Joshi agent-based model for financial markets}
    
    \author[]{Ivan Jericevich}
    \ead{jrciva001@myuct.ac.za}
    \author[]{Murray McKechnie}
    \ead{mckmur003@uct.ac.za}
    \author[]{Tim Gebbie}
    \ead{tim.gebbie@uct.ac.za}
    
    \address[]{Department of Statistical Sciences, University of Cape Town, Rondebosch 7700, South Africa}
    
    \begin{abstract}
        We replicate the contested calibration of the \citet{farmer2002price} agent based model of financial markets using a genetic algorithm and a Nelder-Mead with threshold accepting algorithm following \citet{fabretti2013problem}. The novelty of the Farmer-Joshi model is that the dynamics are driven by trade entry and exit thresholds alone. We recover the known claim that some important stylized facts observed in financial markets cannot be easily found under calibration -- in particular those relating to the auto-correlations in the absolute values of the price fluctuations, and sufficient kurtosis. However, rather than concerns relating to the calibration method, what is novel here is that we extended the Farmer-Joshi model to include agent adaptation using an \citet{brock1998heterogeneous} approach to strategy fitness based on trading strategy profitability. We call this an adaptive Farmer-Joshi model: the model allows trading agents to switch between strategies by favouring strategies that have been more profitable over some period of time determined by a free-parameter fixing the profit monitoring time-horizon. In the adaptive model we are able to calibrate and recover additional stylized facts, despite apparent degeneracy's. This is achieved by combining the interactions of trade entry levels with trade strategy switching. We use this to argue that for low-frequency trading across days, as calibrated to daily sampled data, feed-backs can be accounted for by strategy die-out based on intermediate term profitability; we find that the average trade monitoring horizon is approximately two to three months (or 40 to 60 days) of trading.
    \end{abstract}
\end{frontmatter}

\tableofcontents

\section{Introduction}
This short note is aimed at exploring the calibration and simulation of an intraday ABM as it relates to the more complex topic of intraday ABMs. The discussion here is aimed at trying to find insights into key underlying principles that can drive the relationship between orders in a limit order book. Here the impact of very simple adaption based on intermediate term strategy profitability when trading strategy entry and exit points.

Financial markets exhibit many emergent phenomena, usually attributed to the interactions between the agents that make up the system \cite{heylighen2008complexity}, which causes models that do not take these interactions into account to fail to accurately replicate behaviour observed in the market. Agent-based models (ABMs), meanwhile, demonstrate the ability to produce realistic simulated system dynamics comparable to those observed in empirical investigations \citet{platt2018can}. Successfully calibrating ABMs to financial time series can allow for inference about the factors determining the price behaviour observed in the real world, provided that parameter estimates are sufficiently robust.

A major underlying tenet of the agent-based modelling philosophy is that studying every single element is sufficient to understand the system as a whole \cite{wang2018agent}. While this attempt is able to derive analytical solutions for subsequent analysis, it frequently fails to account successfully for stylized facts, especially in financial markets \cite{wang2018agent}. This is largely because various models in this framework rely heavily on many unrealistic assumptions, such as market clearing, market convergence to equilibrium prices, perfect information, and rationality, whilst ignoring the emergent characteristics of agents’ interactions and their diverse strategies \cite{wang2018agent}.

Overall, ABMs clearly have a number of benefits as a simulation tool to show emergent behaviours, evolution dynamics, and consistent fit with real-market data (for early and more recent reviews of heterogeneous agent models refer to \cite{lebaron2000agent, lebaron2006agent, hommes2006heterogeneous, chiarella2009heterogeneity, chakraborti2011econophysics2, dieci2018heterogeneous, wang2018agent}). ABMs are, however, usually computationally expensive. An additional criticism levelled at agent-based financial markets is that there are too many parameters. Researchers are able not just to move freely through large parameter spaces, but can also change entire internal mechanisms at their discretion in the attempt to fit sets of stylized facts \cite{lebaron2006agent}. Furthermore, unlike analytic models, there are still relatively few general principles that one can confidently apply to the construction of different agent-based market models, and most financial market ABMs assume only a small number of assets \cite{lebaron2006agent}. ABMs also assume that agents operate inductively (use/learn rules and forecasts that have worked well in the past and improve on them). In other words, ABMs do not consider the case where agents operate deductively. Lastly, there is the issue of validation. It is difficult to compare the ``goodness'' of one ABM to another. Such comparisons are usually based on the model's ability to replicate empirical observations - which involves subjective comparisons given the qualitative nature of most stylized facts. It is also usually the case that a unique set of parameters for the calibration of a specific model to a time series does not exist. Without a unique set of parameters allowing us to reproduce the properties of financial time series drawn from a particular market, we cannot argue that introducing and observing changes in the model truly reflects the same changes in the market. Aside from the above points and considering the Farmer-Joshi model specifically, an element missing from its price formation mechanism studied is the risk aversion of the market maker (who is assumed to be risk neutral) which can have profound effects on price formation.

That said, given the uniqueness of different markets, qualitative empirical--versus--simulation comparisons remain the best validation methods. The view point of many market analysts has been and remains an event-based approach in which one attempts to ``explain'' or rationalize a given market movement by relating it to an economic or political event or announcement \cite{cont2001empirical}. From this point of view, one could easily imagine that, since different assets are not necessarily influenced by the same events or information sets, price series obtained from different assets and from different markets will exhibit different properties \cite{cont2001empirical}. Nevertheless, the result of many decades of empirical studies on financial time series indicates that it is the case that if one examines their properties from a statistical point of view, then the seemingly random variations of asset prices do share some quite non-trivial statistical properties \cite{cont2001empirical}. Such properties are known as stylized empirical facts and are frequently used as performance measures for ABMs (for early and more recent extensive reviews on the statistical properties of financial markets refer to \cite{pagan1996econometrics, cont2001empirical, chakraborti2011econophysics1, schmitt2017bimodality}). Stylized facts are thus obtained by taking a common denominator among the properties observed in studies of different markets and instruments. By doing so one gains in generality but tends to lose in precision of the statements one can make about asset returns \cite{cont2001empirical}. Indeed, stylized facts are usually formulated in terms of qualitative properties of asset returns and may not be precise enough to distinguish among different parametric models. Nevertheless, these stylized facts are so constraining that it is not easy to exhibit even an ad hoc stochastic process which possesses the same set of properties and one has to go to great lengths to reproduce them with a model \cite{cont2001empirical}. It should also be noted that to properly validate an agent-based model for financial markets, the model should both be able to replicate the stylized facts of financial markets and have parameters that behave in clear ways and do not have insignificant effects on the resulting behaviour of the simulation. It has, however, been argued that the qualitative comparison of stylized facts is an inadequate form of validation and that they demonstrate an inability to detect parameter degeneracies \cite{platt2016problem}.

The \citet{farmer2002price} model is an inter-day ABM that uses a market maker based method of price formation to study the price dynamics induced by two commonly used financial trading strategies - trend following and value investing - together with state dependent threshold strategies. The use of trade entry and exit levels is very compelling from a practical trading perspective. 

In this implementation of the Farmer-Joshi model, we only consider traders partaking in daily closing auctions. Thus, the model does not face the problem of attempting to model intra-day trader behaviour in a continuous market. While share prices are constantly changing throughout each day, attempting to calibrate ABMs on intra-day prices can cause multiple problems. 

Models considering intra-day price movements can often suffer from parameter degeneracy even if they are able to recreate the stylized facts of the market, suggesting that the parameters no longer carry the meaning they were intended to carry by the model \cite{platt2018can}. This leaves limited ability to make regulatory and structural inferences from the obtained results \cite{platt2018can}. Closing auctions differ from normal intra-day trading in that no sequential order matching is done, with their purpose being to provide a transparent closing price for each share at the end of each day. 

All of the trades that take place at the end of the closing auction execute at the same equilibrium price. Here, going into each closing auction, agents in the model only make use of previous closing prices to determine their actions. As shown in Section \ref{section:Model Formulation}, the Farmer-Joshi model only allows agents to submit market orders, with no allowance for other types of order such as limit orders, stop orders, and stop limit orders \cite{farmer2002price}. With market orders, agents do not attach a price to their order, but instead accept whatever closing price is determined. This means that at the end of the day all orders are always executed in the model. The Farmer-Joshi model therefore does not attempt to include all elements of continuous trading found in real financial markets, instead focusing on a specific aspect of exchange in financial markets. 

For these reasons the \citet{farmer2002price} model is an interesting model for understanding daily trading decisions made from closing auction to closing auction in equity markets, as it attempts to model financial market behaviour without the inclusion of agent adaptation. However, the Farmer-Joshi model is shown to suffer from parameter degeneracy - suggesting that stylized fact centric validation may be insufficient. Nonetheless, as an extension to the Farmer-Joshi model, we consider the case where agents are allowed to switch strategies probabilistically using a \citet{brock1998heterogeneous} approach whereby strategies are favoured according to their profitability over a specified time horizon.

\section{Model Formulation \label{section:Model Formulation}}
\citet{farmer2002price} define a model that considers chartists (trend followers), fundamentalists (value investors), and a risk neutral market maker in order to aggregate the demand of individual agents. The model studies only market orders. During each of $T$ discrete simulation days, the following occurs:
\begin{enumerate}
    \item Each trader agent $i$ observes the most recent prices ($P_t, P_{t-1}, \hdots, P_{t-d}$) and the information $I_t$ and submits order $\omega_{t}^i$ to buy or sell some quantity of an asset to a risk-neutral market maker at day-end.
    \item The market maker then fills the requested orders at a newly determined market price ($P_{t+1}$) based on a closed-form equation aggregating trader agent demands.
\end{enumerate}
Each strategy induces price dynamics that characterize its signal processing properties. Each simulation consists of $N$ traders. At each time step $t$ the $i$th trader sets their order $\omega_t^i$ according to
\begin{equation}
    \omega_t^i = x_t^i - x_{t-1}^i
\end{equation}
where $x_{t}^i = x_t^i(p_t, p_{t-1}, \hdots, I_t)$ is their position at time $t$ determined by their strategy given by function $x_t^i$.

\subsection{Market Maker}
The market maker bases the price formation only on the net order of all traders:
\begin{equation}
    \omega_{t} = \sum_{i = 1}^{N}{\omega_{t}^i}
\end{equation}
The market maker determines the price using the market impact function which relates the net of all orders at a given time to prices. Buying drives the price up, and selling drives it down. Orders, positions and strategies being anonymous motivates basing price formation only on the net order. The function is
\begin{equation} \label{Market impact}
    P_{t+1} = P_t\exp{\frac{\omega_{t}}{\lambda}}
\end{equation}
where $\lambda$ is the liquidity parameter. Letting $p_t = \log{P_t}$ (with $r_t = p_t - p_{t - 1}$ being the log return) and adding a noise term $\zeta_{t+1}$, equation \ref{Market impact} becomes 
\begin{equation}
    p_{t+1} = p_t + \frac{\omega_{t}}{\lambda} + \zeta_{t+1} \hspace{1cm} \zeta_{t+1} \overset{IID}{\sim} N(0, \sigma_{\zeta})
\end{equation}
The random term $\zeta_{t+1}$ can be thought of as corresponding to ``noise traders'' who submit orders at random or as random information that affects the market maker’s price setting decisions \cite{farmer2002price}.

\subsection{Trend Followers}
Trend followers invest based on the belief that price changes have inertia. Hence, we set the $i$th trend follower's position at time $t$ according to
\begin{equation}
    x^i_{t+1} = c^i \cdot \text{sign} (p_t - p_{t-d^i}) \hspace{1cm} d^i \sim U(d_{min}, d_{max})
\end{equation}
where $c^i > 0$ is a constant proportional to trading capital and $d^i$ is the time lag of the $i$th agent. The model assumes chartists care only about the direction of the change from lagged price to current price and not the magnitude. Chartists amplify noise in prices by reinforcing price movements, leading to extra volatility and irrational valuations of the security which make information aggregation and dissemination more difficult.

\subsection{Value Investors}
Value investors believe their perceived value may not be fully reflected in the current price, and that the price will move towards their perceived value. Hence, their decisions are based on price deviations from a perceived fundamental asset value. The $i$th value investor's position at time $t$ is set according to
\begin{align} \label{Value updating equation}
\begin{split}
    x_{t+1}^i &= c^i \cdot \text{sign} (v^i_t - p_t) \hspace{5mm} v^i_1 \sim U(v_{min}, v_{max}) \\
    v^i_{t+1} &= v^i_t + \eta_{t+1} \hspace{5mm} \eta_{t+1} \overset{IID}{\sim} N(\mu_\eta, \sigma_\eta)
\end{split}
\end{align}
where $v^i_t$ is the log of an investor's value perception at time $t$. Similar to chartists, we assume fundamentalists care only about the direction of the change from the current price to the perceived value. For the purposes of this paper it does not matter how individual agents form their opinions about value. We take the estimated value as an exogenous input, and focus on the response of prices to changes in it \cite{farmer2002price}.

\subsection{State-Dependent Threshold Strategies}
A concern with simple position-based value/trend strategies is that trades are made whenever the mispricing changes, but this would lead to excessive transaction costs in a real-world setting. To ameliorate this problem, a threshold is used for entering and exiting positions. As a result, not all fundamentalist/chartist trader agents are active in all simulation steps, but rather activate only when a certain threshold of mispricing ($m^i_t = p_t - v^i_t$ for fundamentalists or $m^i_t = p_{t - d^i} - p_t$ for chartists) is realized. The model assumes each trader has an associated position entry threshold, $T^i$ where $T^i \sim U(T_{min}, T_{max})$ for each trader in the simulation, and an associated position exit threshold, $\tau^i$ where $\tau^i \sim U(\tau_{min}, \tau_{max})$. When $m^i_t$ crosses below $-T^i$, the trader enters long position $c^i$ and only exits it when $m^i_t$ crosses back above $-\tau^i$. Similarly, when $m^i_t$ crosses above $T^i$, the trader enters short position $-c^i$ and only exits it when $m^i_t$ crosses below $\tau^i$. Finally, we set $c^i = a(T^i - \tau^i)$, where the scale parameter for capital assignment $a$ is a positive constant.

\subsection{Agent Adaptation}
The above formulation completely characterises the standard \citet{farmer2002price} model. The concept of changes in the number of chartists and fundamentalists is driven by the Friedman hypothesis: ``irrational agents will lose money and will be driven out the market by rational agents'' in a predator-prey type fashion \cite{friedman1953essays}. An important question in heterogeneous agents modelling is whether ``irrational'' traders can survive in the market, or whether they would lose money and be driven out of the market by rational investors, who would trade against them and drive prices back to fundamentals, as argued by Friedman \cite{friedman1953essays}. 

On the other hand, a strong argument can be made for refraining from allowing for agent adaptation in ABMs. For example, it would be desirable if the original Farmer-Joshi model were able to replicate the stylized facts, as the authors argue that traders do not constantly consider different strategies and change between them as is often the case in adaptive agent models \cite{farmer2002price}. Ideally Farmer and Joshi wish to be able to replicate observed price behaviour from the internal mechanisms of agents' strategies and the different ways in which the two strategies are activated \cite{farmer2002price}. By including adaptive agents in the model, we are adding an explicit model definition in order to better replicate the stylized facts but at the expense of possibly reducing the realism of the agent behaviour, even if such behaviour could be argued to be more rational than simply sticking to a predefined strategy.

The idea behind the adaptive model is that agents revise their ``beliefs'' in each period in a boundedly rational way, according to a ``fitness measure'' - past realized profits \cite{brock1998heterogeneous}. Trend trading is one of the most important factors leading to excess volatility. So, we include a system which allows for the inclusion of more trend followers during periods of the simulation. By the nature of their strategies, chartists induce positive short-term autocorrelations and value investors induce negative short-term autocorrelations.

The number of traders in the system is fixed at $N$, with the number of available strategies being $2N$ (two for each trader). We denote the $i$th agent's available fundamentalist and chartist strategy positions at time $t$ by $x_t^{f, i}$ and $x_t^{c, i}$ respectively. Traders track profits over the $H$ most recent days according to 
\begin{equation}
    \pi^{s, i}_t = \sum_{k = t - H + 1}^t (x^{s, i}_{k - 1}(p_k - p_{k - 1}))
\end{equation}
where $s$ denotes the strategy adopted, and $\pi^{s, i}_t$ is the $i$th agent's profit earned from their strategy at time $t$. In order to analyse the long time dynamics of agent adaptation, it is useful to get rid of the discrete characteristics of the game, such as the ``always play the best strategy'' rule. Instead, we use a probabilistic strategy selection rule, as introduced by \citet{cavagna1999thermal}, that favours well performing strategies. We impose that agent $i$ adopts their strategy $s$ on day $t$ with probability given by
\begin{equation}
    \phi^{c, i}_t = \frac{e^{\pi^{c, i}_t/\Gamma}}{e^{\pi^{c, i}_t/\Gamma} + e^{\pi^{f, i}_t/\Gamma}} \hspace{1cm} \text{and} \hspace{1cm} \phi^{f, i}_t = 1 - \phi^{c, i}_t
\end{equation}
where $\Gamma$ is a positive constant for intensity of switching and $\phi^{s, i}_t$ is the $i$th agent's probability of adopting strategy $s \in$ \{fundamentalist ($f$), chartist ($c$)\} at time $t$. The intensity of switching parameter determines how sensitive the traders are to differences in profitability across trading strategies. Agents reassess their strategy each day, but only change their position if the relevant entry or exit threshold is met.

The strategies mentioned are only a few of the strategies actually used in real markets. But they are known to be widely used, and understanding their influence on prices provides a starting point for more realistic behavioural models.

\section{Model Calibration}
\begin{figure}[H]
    \centering
    \begin{subfigure}[t]{0.7\linewidth}
        \includegraphics[trim = 65 55 65 70, clip, width = \linewidth]{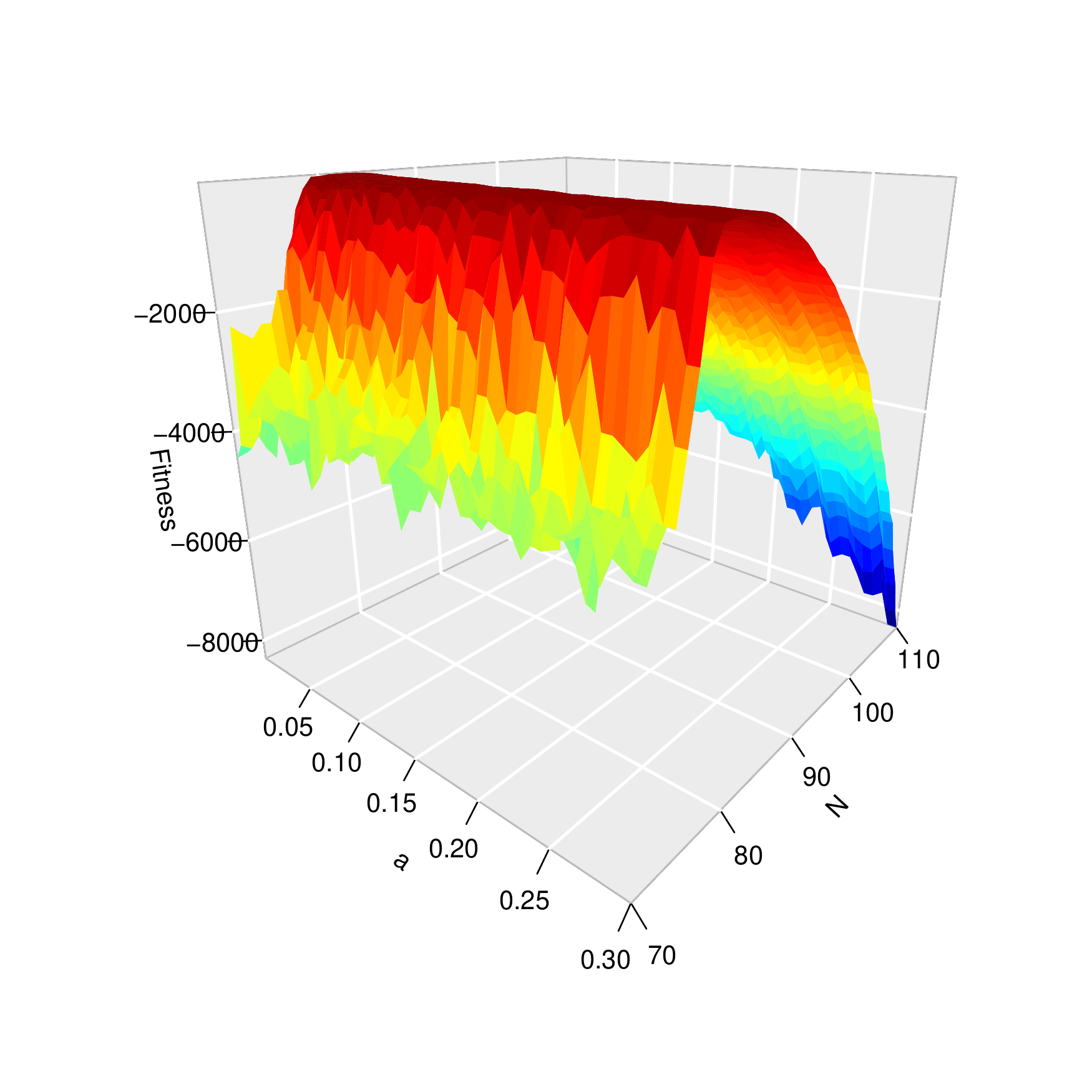}
        \caption{$a, N$}
    \end{subfigure}
    \begin{subfigure}[t]{0.7\linewidth}
        \includegraphics[trim = 62 55 65 70, clip, width = \linewidth]{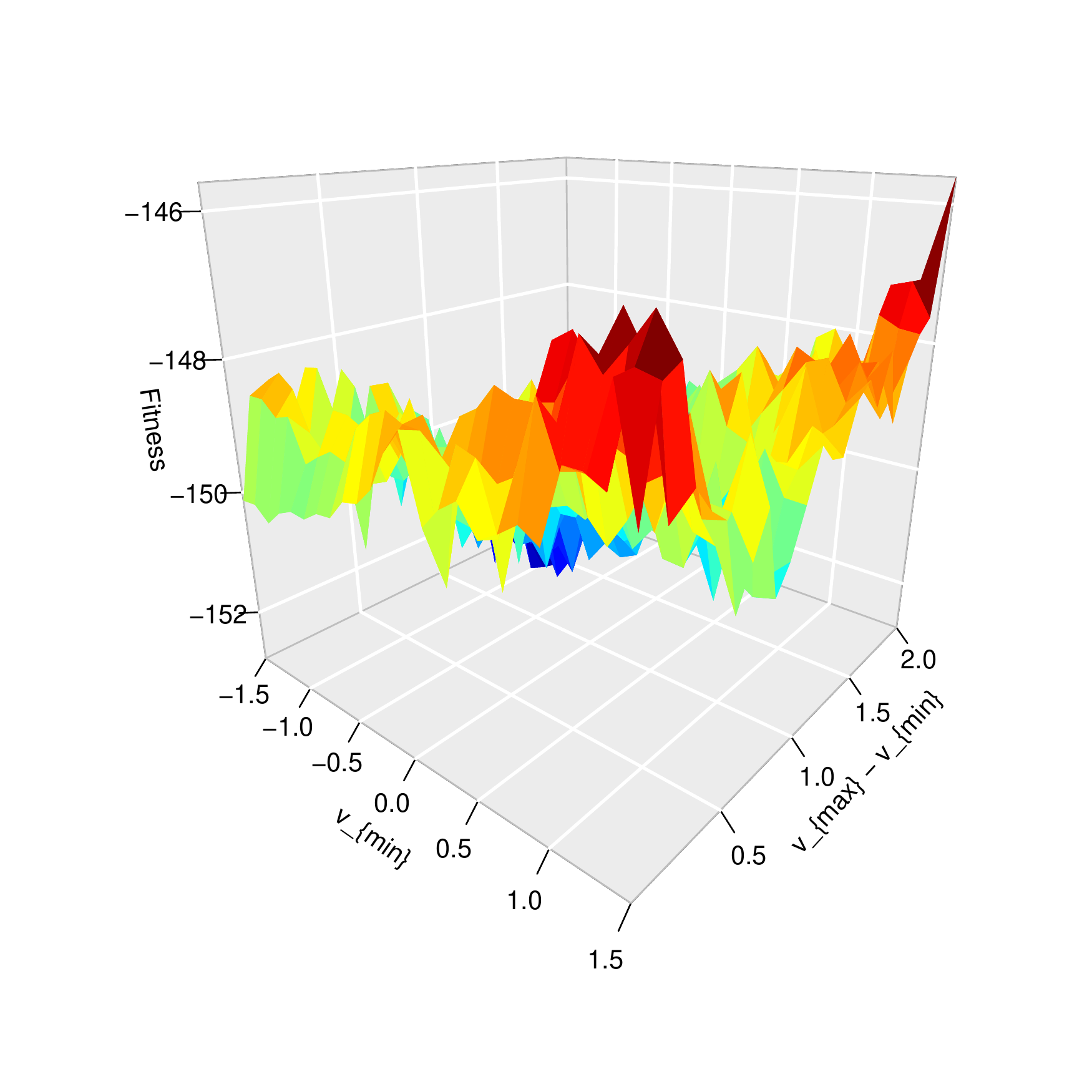}
        \caption{$v_{min}, v_{max} - v_{min}$}
    \end{subfigure}
    \caption{Marginal objective surfaces \label{fig:Objective-Surfaces}}
\end{figure}

We restrict ourselves to an objective function based frequentist approach to model calibration, but note that such practices have been criticised for their rigour \cite{platt2020comparison, fabretti2013problem}. Their under-performance is argued to be the result of the objective functions lacking smoothness and having many local optimums, making discovery of the global optimum difficult \cite{gilli2003global} (see figure \ref{fig:Objective-Surfaces}b). Such an objective surface indicates parameter degeneracies - when a parameter has no clear unique optimum after many independent calibration attempts. For this reason, the use of heuristic methods can be preferable for ABMs, as a heuristic method can be better at obtaining an approximation of the global optimum. Threshold accepting is a heuristic search method which can be used in conjunction with the Nelder-Mead simplex algorithm to calibrate ABMs. Gilli and Winker \cite{gilli2003global} present a global optimisation heuristic using this Nelder-Mead with threshold accepting technique.

Any solution for the optimum is only approximate due to the stochastic nature of ABMs \cite{fabretti2013problem}, so we use heuristic methods for calibration. A good objective function can be constructed by making use of a set of $k$ moments that best describe the statistical features of financial data \cite{winker2007objective}. Calibration is then a matter of matching empirical moments with those of simulated data - the method of moments. The objective is thus the minimization of estimation errors of predetermined moments:
\begin{equation}
    \min \limits_{\bm{\theta} \in \bm{\Theta}} f(\bm{\theta})
\end{equation}
where $\bm{\theta}$ is the vector of parameters and $\bm{\Theta}$ the space of feasible parameters. Since some properties may be more important for replicating stylized facts, we must determine how each of the chosen moments and statistics are weighted. It cannot be assumed that the moments are distributed independently from each other, which motivates the use of a weight matrix that considers the joint distribution of moments \cite{winker2007objective}. Denoting by $\bm{m^e} = [m_1^e, \hdots, m_k^e]^{\prime}$ and $[\bm{m^s} \mid \bm{\theta}] = [m_1^s, \hdots, m_k^s]^{\prime}$ the vectors of empirical moments and statistics of real and simulated data respectively, and using the empirical moments $\bm{m^e}$ to estimate the true moments, we define the estimation error as the mean deviation of simulated from empirical moments (we require that $E[[\bm{m^s} \mid \bm{\theta}]] = \bm{m} \implies E[[\bm{m^s} \mid \bm{\theta}] - \bm{m}] = 0$) \cite{heij2004econometric}:
\begin{equation}
    G(\bm{\theta}) = \frac{1}{I} \sum_{i = 1}^{I}{(\bm{m_i^e} - [\bm{m_i^s} \mid \bm{\theta}])}
\end{equation}
where $I$ is the number of simulations used. A high value of $I$ is preferable as it reduces objective function variance, but this costs computational time. Denoting by $\bm{W}$ the $k \times k$ weight matrix of moments and statistics, the objective function is the sum of squares of the deviations of simulated from empirical moments:
\begin{equation}
    f(\bm{\theta}) = G(\bm{\theta})^{\prime}\bm{W}G(\bm{\theta})
    \label{fitness equation}
\end{equation}
According to \citet{heij2004econometric}, the matrix $\bm{W}$ is given by the inverse of the covariance matrix of the distribution of moments ($Var^{-1}[\bm{m^e}]$). This weight matrix takes the uncertainty of estimation associated with $\bm{m^e}$ into account by assigning larger weights to moments associated with lower uncertainty.

$\bm{W}$ is estimated by applying a moving block bootstrap to the time series with 100-day blocks \cite{fabretti2013problem, platt2016problem}. This method is used over case resampling to ensure the auto-correlations present in financial market time series are not ignored. It is important to choose the block size such that a balance is struck between the preservation of auto-correlations and the estimation accuracy of the variance in moments and statistics.

The choice of moments and statistics should be robust enough to reflect the properties of financial data and flexible enough to discriminate between different models \cite{winker2007objective}. We select the following $k = 9$ moments and statistics in the objective function based on \citet{winker2007objective}: mean, standard deviation, excess kurtosis, Komogorov-Smirnov (K-S) statistic; simplified Hurst exponent, Geweke and Porter-Hudak (GPH) estimator, Augmented Dickey-Fuller (ADF) statistic; the sum of the two GARCH(1, 1) parameters and the average of the Hill estimator on the right tail of the distribution of returns from the $90^{th}$ to the $95^{th}$ percentile\footnote{The mean, standard deviation, kurtosis, Hill estimator and the K-S statistics are used to represent the overall shape of the data distribution; Hurst exponent represents the scaling properties of the log returns; GPH estimator provides a measure of the long-range dependence of the absolute log returns; ADF statistic is a measure of the extent of the random walk property of log returns; sum of GARCH parameters is a measure of short range dependence; Hill estimator is a measure of the fat tails on the right-hand side of the return distribution}.

Given the above objective function, the parameters optimised are given in table \ref{table:Parameters}\footnote{$N_f, N_c$ - number of traders of each type; $\lambda$ - liquidity; $a$ - scale parameter for capital assignment; $d_{min}$, $d_{max}$ - minimum and maximum time delay (chartists); $\mu_{\eta}$, $\sigma_{\eta}$ - mean and s.d. of noise process in $v_t$; $\sigma_{\zeta}$ - noise variance in price setting function; $T_{min}$, $T_{max}$ - Entering position thresholds; $\tau_{min}$, $\tau_{max}$ - Exiting position thresholds; $v_{min}$, $v_{max}$ - Log perceived value offset range (fundamentalists); $\Gamma$ - Switching intensity; $H$ - Profit tracking time horizon}. We use a Genetic algorithm (GA) and a Nelder-Mead with threshold accepting (NMTA) algorithm for their ability to explore the global solution space efficiently. The NMTA algorithm combines the standard Nelder-Mead method \cite{nelder1965simplex} with an occasional threshold accepting step that shifts the entire simplex provided the fitness does not drop by a predefined threshold - which helps the algorithm escape sub-optimal local minima \cite{gilli2003global}.

\section{Results and Analysis}
The data used to calibrate the Farmer-Joshi model was Anglo American daily sampled closing price data from 01/01/2005 to 29/04/2016 obtained from the \href{https://data.mendeley.com/datasets/3nbgc4cygk/1}{Mendeley} website \cite{gant2019jse}.

\subsection{Calibration}
In table \ref{table:Parameters} the point estimates are the estimates of the parameters obtained by each optimisation algorithm that resulted in the best fitness. The confidence intervals indicate that many of the parameters can vary substantially from one calibration to another, even with the fitness obtained being very similar. This indicates the existence of many local optima, each with a similar ability to replicate the moments and statistics found in the actual data. These large confidence intervals unfortunately limit the explanatory power of the model. Even with more advanced calibration methods it seems unlikely that these types of degeneracy's can be overcome in a useful way. 

The value for $\Gamma$ obtained by the GA for the adaptive model is larger than that of the NMTA algorithm. Given that the NMTA algorithm was less able to replicate the stylized facts (figure \ref{fig:AFJ-Results}), the larger value for $\Gamma$ (intensity of switching, thereby allowing for more periods where chartist activity dominates) appears preferable for the replication of stylized facts.
\begin{table*}
    \small
    \setlength{\tabcolsep}{1pt}
    \centering
    \begin{tabular}{lcccc|cccc} \toprule
        {} & \multicolumn{4}{c}{Standard Farmer-Joshi} & \multicolumn{4}{c}{Adaptive Farmer-Joshi} \\ \cmidrule{2-5} \cmidrule{6-9}
        {$\theta$} & {$\theta_{GA}$} & {$\theta^{95\%}_{GA}$} & {$\theta_{NMTA}$} & {$\theta^{95\%}_{NMTA}$} & {$\theta_{GA}$} & {$\theta^{95\%}_{GA}$} & {$\theta_{NMTA}$} & {$\theta^{95\%}_{NMTA}$} \\ \midrule
        $N_f, N_c$ & \DTLfetch{Parameters (standard)}{Parameter}{N}{GA} & [\DTLfetch{Parameters (standard)}{Parameter}{N}{Lower_GA}; \DTLfetch{Parameters (standard)}{Parameter}{N}{Upper_GA}] & \DTLfetch{Parameters (standard)}{Parameter}{N}{NMTA} & [\DTLfetch{Parameters (standard)}{Parameter}{N}{Lower_NMTA}; \DTLfetch{Parameters (standard)}{Parameter}{N}{Upper_NMTA}] & \DTLfetch{Parameters (adaptive)}{Parameter}{N}{GA} & [\DTLfetch{Parameters (adaptive)}{Parameter}{N}{Lower_GA}; \DTLfetch{Parameters (adaptive)}{Parameter}{N}{Upper_GA}] & \DTLfetch{Parameters (adaptive)}{Parameter}{N}{NMTA} & [\DTLfetch{Parameters (adaptive)}{Parameter}{N}{Lower_NMTA}; \DTLfetch{Parameters (adaptive)}{Parameter}{N}{Upper_NMTA}] \\
        $\lambda$ & \DTLfetch{Parameters (standard)}{Parameter}{lambda}{GA} & [\DTLfetch{Parameters (standard)}{Parameter}{lambda}{Lower_GA}; \DTLfetch{Parameters (standard)}{Parameter}{lambda}{Upper_GA}] & \DTLfetch{Parameters (standard)}{Parameter}{lambda}{NMTA} & [\DTLfetch{Parameters (standard)}{Parameter}{lambda}{Lower_NMTA}; \DTLfetch{Parameters (standard)}{Parameter}{lambda}{Upper_NMTA}] & \DTLfetch{Parameters (adaptive)}{Parameter}{lambda}{GA} & [\DTLfetch{Parameters (adaptive)}{Parameter}{lambda}{Lower_GA}; \DTLfetch{Parameters (adaptive)}{Parameter}{lambda}{Upper_GA}] & \DTLfetch{Parameters (adaptive)}{Parameter}{lambda}{NMTA} & [\DTLfetch{Parameters (adaptive)}{Parameter}{lambda}{Lower_NMTA}; \DTLfetch{Parameters (adaptive)}{Parameter}{lambda}{Upper_NMTA}] \\
        $a$ & \DTLfetch{Parameters (standard)}{Parameter}{a}{GA} & [\DTLfetch{Parameters (standard)}{Parameter}{a}{Lower_GA}; \DTLfetch{Parameters (standard)}{Parameter}{a}{Upper_GA}] & \DTLfetch{Parameters (standard)}{Parameter}{a}{NMTA} & [\DTLfetch{Parameters (standard)}{Parameter}{a}{Lower_NMTA}; \DTLfetch{Parameters (standard)}{Parameter}{a}{Upper_NMTA}] & \DTLfetch{Parameters (adaptive)}{Parameter}{a}{GA} & [\DTLfetch{Parameters (adaptive)}{Parameter}{a}{Lower_GA}; \DTLfetch{Parameters (adaptive)}{Parameter}{a}{Upper_GA}] & \DTLfetch{Parameters (adaptive)}{Parameter}{a}{NMTA} & [\DTLfetch{Parameters (adaptive)}{Parameter}{a}{Lower_NMTA}; \DTLfetch{Parameters (adaptive)}{Parameter}{a}{Upper_NMTA}] \\
        $d_{max}$ & \DTLfetch{Parameters (standard)}{Parameter}{d_max}{GA} & [\DTLfetch{Parameters (standard)}{Parameter}{d_max}{Lower_GA}; \DTLfetch{Parameters (standard)}{Parameter}{d_max}{Upper_GA}] & \DTLfetch{Parameters (standard)}{Parameter}{d_max}{NMTA} & [\DTLfetch{Parameters (standard)}{Parameter}{d_max}{Lower_NMTA}; \DTLfetch{Parameters (standard)}{Parameter}{d_max}{Upper_NMTA}] & \DTLfetch{Parameters (adaptive)}{Parameter}{d_max}{GA} & [\DTLfetch{Parameters (adaptive)}{Parameter}{d_max}{Lower_GA}; \DTLfetch{Parameters (adaptive)}{Parameter}{d_max}{Upper_GA}] & \DTLfetch{Parameters (adaptive)}{Parameter}{d_max}{NMTA} & [\DTLfetch{Parameters (adaptive)}{Parameter}{d_max}{Lower_NMTA}; \DTLfetch{Parameters (adaptive)}{Parameter}{d_max}{Upper_NMTA}] \\
        $d_{min}$ & \DTLfetch{Parameters (standard)}{Parameter}{d_min}{GA} & [\DTLfetch{Parameters (standard)}{Parameter}{d_min}{Lower_GA}; \DTLfetch{Parameters (standard)}{Parameter}{d_min}{Upper_GA}] & \DTLfetch{Parameters (standard)}{Parameter}{d_min}{NMTA} & [\DTLfetch{Parameters (standard)}{Parameter}{d_min}{Lower_NMTA}; \DTLfetch{Parameters (standard)}{Parameter}{d_min}{Upper_NMTA}] & \DTLfetch{Parameters (adaptive)}{Parameter}{d_min}{GA} & [\DTLfetch{Parameters (adaptive)}{Parameter}{d_min}{Lower_GA}; \DTLfetch{Parameters (adaptive)}{Parameter}{d_min}{Upper_GA}] & \DTLfetch{Parameters (adaptive)}{Parameter}{d_min}{NMTA} & [\DTLfetch{Parameters (adaptive)}{Parameter}{d_min}{Lower_NMTA}; \DTLfetch{Parameters (adaptive)}{Parameter}{d_min}{Upper_NMTA}] \\
        $\mu_{\eta}$ & \DTLfetch{Parameters (standard)}{Parameter}{mu_eta}{GA} & [\DTLfetch{Parameters (standard)}{Parameter}{mu_eta}{Lower_GA}; \DTLfetch{Parameters (standard)}{Parameter}{mu_eta}{Upper_GA}] & \DTLfetch{Parameters (standard)}{Parameter}{mu_eta}{NMTA} & [\DTLfetch{Parameters (standard)}{Parameter}{mu_eta}{Lower_NMTA}; \DTLfetch{Parameters (standard)}{Parameter}{mu_eta}{Upper_NMTA}] & \DTLfetch{Parameters (adaptive)}{Parameter}{mu_eta}{GA} & [\DTLfetch{Parameters (adaptive)}{Parameter}{mu_eta}{Lower_GA}; \DTLfetch{Parameters (adaptive)}{Parameter}{mu_eta}{Upper_GA}] & \DTLfetch{Parameters (adaptive)}{Parameter}{mu_eta}{NMTA} & [\DTLfetch{Parameters (adaptive)}{Parameter}{mu_eta}{Lower_NMTA}; \DTLfetch{Parameters (adaptive)}{Parameter}{mu_eta}{Upper_NMTA}] \\
        $\sigma_{\eta}$ & \DTLfetch{Parameters (standard)}{Parameter}{sigma_eta}{GA} & [\DTLfetch{Parameters (standard)}{Parameter}{sigma_eta}{Lower_GA}; \DTLfetch{Parameters (standard)}{Parameter}{sigma_eta}{Upper_GA}] & \DTLfetch{Parameters (standard)}{Parameter}{sigma_eta}{NMTA} & [\DTLfetch{Parameters (standard)}{Parameter}{sigma_eta}{Lower_NMTA}; \DTLfetch{Parameters (standard)}{Parameter}{sigma_eta}{Upper_NMTA}] & \DTLfetch{Parameters (adaptive)}{Parameter}{sigma_eta}{GA} & [\DTLfetch{Parameters (adaptive)}{Parameter}{sigma_eta}{Lower_GA}; \DTLfetch{Parameters (adaptive)}{Parameter}{sigma_eta}{Upper_GA}] & \DTLfetch{Parameters (adaptive)}{Parameter}{sigma_eta}{NMTA} & [\DTLfetch{Parameters (adaptive)}{Parameter}{sigma_eta}{Lower_NMTA}; \DTLfetch{Parameters (adaptive)}{Parameter}{sigma_eta}{Upper_NMTA}] \\
        $\sigma_{\zeta}$ & \DTLfetch{Parameters (standard)}{Parameter}{sigma_zeta}{GA} & [\DTLfetch{Parameters (standard)}{Parameter}{sigma_zeta}{Lower_GA}; \DTLfetch{Parameters (standard)}{Parameter}{sigma_zeta}{Upper_GA}] & \DTLfetch{Parameters (standard)}{Parameter}{sigma_zeta}{NMTA} & [\DTLfetch{Parameters (standard)}{Parameter}{sigma_zeta}{Lower_NMTA}; \DTLfetch{Parameters (standard)}{Parameter}{sigma_zeta}{Upper_NMTA}] & \DTLfetch{Parameters (adaptive)}{Parameter}{sigma_zeta}{GA} & [\DTLfetch{Parameters (adaptive)}{Parameter}{sigma_zeta}{Lower_GA}; \DTLfetch{Parameters (adaptive)}{Parameter}{sigma_zeta}{Upper_GA}] & \DTLfetch{Parameters (adaptive)}{Parameter}{sigma_zeta}{NMTA} & [\DTLfetch{Parameters (adaptive)}{Parameter}{sigma_zeta}{Lower_NMTA}; \DTLfetch{Parameters (adaptive)}{Parameter}{sigma_zeta}{Upper_NMTA}] \\
        $T_{max}$ & \DTLfetch{Parameters (standard)}{Parameter}{T_max}{GA} & [\DTLfetch{Parameters (standard)}{Parameter}{T_max}{Lower_GA}; \DTLfetch{Parameters (standard)}{Parameter}{T_max}{Upper_GA}] & \DTLfetch{Parameters (standard)}{Parameter}{T_max}{NMTA} & [\DTLfetch{Parameters (standard)}{Parameter}{T_max}{Lower_NMTA}; \DTLfetch{Parameters (standard)}{Parameter}{T_max}{Upper_NMTA}] & \DTLfetch{Parameters (adaptive)}{Parameter}{T_max}{GA} & [\DTLfetch{Parameters (adaptive)}{Parameter}{T_max}{Lower_GA}; \DTLfetch{Parameters (adaptive)}{Parameter}{T_max}{Upper_GA}] & \DTLfetch{Parameters (adaptive)}{Parameter}{T_max}{NMTA} & [\DTLfetch{Parameters (adaptive)}{Parameter}{T_max}{Lower_NMTA}; \DTLfetch{Parameters (adaptive)}{Parameter}{T_max}{Upper_NMTA}] \\
        $T_{min}$ & \DTLfetch{Parameters (standard)}{Parameter}{T_min}{GA} & [\DTLfetch{Parameters (standard)}{Parameter}{T_min}{Lower_GA}; \DTLfetch{Parameters (standard)}{Parameter}{T_min}{Upper_GA}] & \DTLfetch{Parameters (standard)}{Parameter}{T_min}{NMTA} & [\DTLfetch{Parameters (standard)}{Parameter}{T_min}{Lower_NMTA}; \DTLfetch{Parameters (standard)}{Parameter}{T_min}{Upper_NMTA}] & \DTLfetch{Parameters (adaptive)}{Parameter}{T_min}{GA} & [\DTLfetch{Parameters (adaptive)}{Parameter}{T_min}{Lower_GA}; \DTLfetch{Parameters (adaptive)}{Parameter}{T_min}{Upper_GA}] & \DTLfetch{Parameters (adaptive)}{Parameter}{T_min}{NMTA} & [\DTLfetch{Parameters (adaptive)}{Parameter}{T_min}{Lower_NMTA}; \DTLfetch{Parameters (adaptive)}{Parameter}{T_min}{Upper_NMTA}] \\
        $\tau_{min}$ & \DTLfetch{Parameters (standard)}{Parameter}{tau_min}{GA} & [\DTLfetch{Parameters (standard)}{Parameter}{tau_min}{Lower_GA}; \DTLfetch{Parameters (standard)}{Parameter}{tau_min}{Upper_GA}] & \DTLfetch{Parameters (standard)}{Parameter}{tau_min}{NMTA} & [\DTLfetch{Parameters (standard)}{Parameter}{tau_min}{Lower_NMTA}; \DTLfetch{Parameters (standard)}{Parameter}{tau_min}{Upper_NMTA}] & \DTLfetch{Parameters (adaptive)}{Parameter}{tau_min}{GA} & [\DTLfetch{Parameters (adaptive)}{Parameter}{tau_min}{Lower_GA}; \DTLfetch{Parameters (adaptive)}{Parameter}{tau_min}{Upper_GA}] & \DTLfetch{Parameters (adaptive)}{Parameter}{tau_min}{NMTA} & [\DTLfetch{Parameters (adaptive)}{Parameter}{tau_min}{Lower_NMTA}; \DTLfetch{Parameters (adaptive)}{Parameter}{tau_min}{Upper_NMTA}] \\
        $\tau_{max}$ & \DTLfetch{Parameters (standard)}{Parameter}{tau_max}{GA} & [\DTLfetch{Parameters (standard)}{Parameter}{tau_max}{Lower_GA}; \DTLfetch{Parameters (standard)}{Parameter}{tau_max}{Upper_GA}] & \DTLfetch{Parameters (standard)}{Parameter}{tau_max}{NMTA} & [\DTLfetch{Parameters (standard)}{Parameter}{tau_max}{Lower_NMTA}; \DTLfetch{Parameters (standard)}{Parameter}{tau_max}{Upper_NMTA}] & \DTLfetch{Parameters (adaptive)}{Parameter}{tau_max}{GA} & [\DTLfetch{Parameters (adaptive)}{Parameter}{tau_max}{Lower_GA}; \DTLfetch{Parameters (adaptive)}{Parameter}{tau_max}{Upper_GA}] & \DTLfetch{Parameters (adaptive)}{Parameter}{tau_max}{NMTA} & [\DTLfetch{Parameters (adaptive)}{Parameter}{tau_max}{Lower_NMTA}; \DTLfetch{Parameters (adaptive)}{Parameter}{tau_max}{Upper_NMTA}] \\
        $v_{max}$ & \DTLfetch{Parameters (standard)}{Parameter}{v_max}{GA} & [\DTLfetch{Parameters (standard)}{Parameter}{v_max}{Lower_GA}; \DTLfetch{Parameters (standard)}{Parameter}{v_max}{Upper_GA}] & \DTLfetch{Parameters (standard)}{Parameter}{v_max}{NMTA} & [\DTLfetch{Parameters (standard)}{Parameter}{v_max}{Lower_NMTA}; \DTLfetch{Parameters (standard)}{Parameter}{v_max}{Upper_NMTA}] & \DTLfetch{Parameters (adaptive)}{Parameter}{v_max}{GA} & [\DTLfetch{Parameters (adaptive)}{Parameter}{v_max}{Lower_GA}; \DTLfetch{Parameters (adaptive)}{Parameter}{v_max}{Upper_GA}] & \DTLfetch{Parameters (adaptive)}{Parameter}{v_max}{NMTA} & [\DTLfetch{Parameters (adaptive)}{Parameter}{v_max}{Lower_NMTA}; \DTLfetch{Parameters (adaptive)}{Parameter}{v_max}{Upper_NMTA}] \\
        $v_{min}$ & \DTLfetch{Parameters (standard)}{Parameter}{v_min}{GA} & [\DTLfetch{Parameters (standard)}{Parameter}{v_min}{Lower_GA}; \DTLfetch{Parameters (standard)}{Parameter}{v_min}{Upper_GA}] & \DTLfetch{Parameters (standard)}{Parameter}{v_min}{NMTA} & [\DTLfetch{Parameters (standard)}{Parameter}{v_min}{Lower_NMTA}; \DTLfetch{Parameters (standard)}{Parameter}{v_min}{Upper_NMTA}] & \DTLfetch{Parameters (adaptive)}{Parameter}{v_min}{GA} & [\DTLfetch{Parameters (adaptive)}{Parameter}{v_min}{Lower_GA}; \DTLfetch{Parameters (adaptive)}{Parameter}{v_min}{Upper_GA}] & \DTLfetch{Parameters (adaptive)}{Parameter}{v_min}{NMTA} & [\DTLfetch{Parameters (adaptive)}{Parameter}{v_min}{Lower_NMTA}; \DTLfetch{Parameters (adaptive)}{Parameter}{v_min}{Upper_NMTA}] \\
        $\Gamma$ & - & - & - & - & \DTLfetch{Parameters (adaptive)}{Parameter}{Gamma}{GA} & [\DTLfetch{Parameters (adaptive)}{Parameter}{Gamma}{Lower_GA}; \DTLfetch{Parameters (adaptive)}{Parameter}{Gamma}{Upper_GA}] & \DTLfetch{Parameters (adaptive)}{Parameter}{Gamma}{NMTA} & [\DTLfetch{Parameters (adaptive)}{Parameter}{Gamma}{Lower_NMTA}; \DTLfetch{Parameters (adaptive)}{Parameter}{Gamma}{Upper_NMTA}] \\
        $H$ & - & - & - & - & \DTLfetch{Parameters (adaptive)}{Parameter}{H}{GA} & [\DTLfetch{Parameters (adaptive)}{Parameter}{H}{Lower_GA}; \DTLfetch{Parameters (adaptive)}{Parameter}{H}{Upper_GA}] & \DTLfetch{Parameters (adaptive)}{Parameter}{H}{NMTA} & [\DTLfetch{Parameters (adaptive)}{Parameter}{H}{Lower_NMTA}; \DTLfetch{Parameters (adaptive)}{Parameter}{H}{Upper_NMTA}] \\
        Fitness & \DTLfetch{Parameters (standard)}{Parameter}{Fitness}{GA} & [\DTLfetch{Parameters (standard)}{Parameter}{Fitness}{Lower_GA}; \DTLfetch{Parameters (standard)}{Parameter}{Fitness}{Upper_GA}] & \DTLfetch{Parameters (standard)}{Parameter}{Fitness}{NMTA} & [\DTLfetch{Parameters (standard)}{Parameter}{Fitness}{Lower_NMTA}; \DTLfetch{Parameters (standard)}{Parameter}{Fitness}{Upper_NMTA}] & \DTLfetch{Parameters (adaptive)}{Parameter}{Fitness}{GA} & [\DTLfetch{Parameters (adaptive)}{Parameter}{Fitness}{Lower_GA}; \DTLfetch{Parameters (adaptive)}{Parameter}{Fitness}{Upper_GA}] & \DTLfetch{Parameters (adaptive)}{Parameter}{Fitness}{NMTA} & [\DTLfetch{Parameters (adaptive)}{Parameter}{Fitness}{Lower_NMTA}; \DTLfetch{Parameters (adaptive)}{Parameter}{Fitness}{Upper_NMTA}] \\ \bottomrule
    \end{tabular}
    \caption{\label{table:Parameters} Estimates of parameters using GA and NMTA algorithms.}
\end{table*}
Table \ref{table:Moments and Statistics} provides confidence intervals for the moments and statistics of simulated data for both models and calibration methods along with the empirical values $m^e$. For the standard model, we note that the optimal parameters from both calibration methods struggle in particular to replicate the GPH estimator. For many of the moments, we observe wide confidence intervals, indicating a lack of consistency across simulations.

For the adaptive model, the moments of simulated log returns more closely match the actual returns, and the confidence intervals mostly encapsulate the empirical values. The GA calibration results in greater excess kurtosis than the NMTA calibration, implying more leptokurtic and fat-tailed log returns.
\begin{sidewaystable}[p]
    \small
    \setlength{\tabcolsep}{2pt}
    \centering
    \begin{tabular}{lcccc|cccc|c} \toprule
        {} & \multicolumn{4}{c}{Standard Farmer-Joshi} & \multicolumn{4}{c}{Adaptive Farmer-Joshi} & {Empirical} \\ \cmidrule{2-5} \cmidrule{6-10}
        {Moments and statistics} & {$m^s|\theta_{GA}$} & {$\text{CI}^{95\%}_{GA}$} & {$m^s|\theta_{NMTA}$} & {$\text{CI}^{95\%}_{NMTA}$} & {$m^s|\theta_{GA}$} & {$\text{CI}^{95\%}_{GA}$} & {$m^s|\theta_{NMTA}$} & {$\text{CI}^{95\%}_{NMTA}$} & {$m^e$} \\ \midrule
        Mean & \DTLfetch{Moments (standard)}{Moment}{mean}{GA} & [\DTLfetch{Moments (standard)}{Moment}{mean}{Lower_GA}; \DTLfetch{Moments (standard)}{Moment}{mean}{Upper_GA}] & \DTLfetch{Moments (standard)}{Moment}{mean}{NMTA} & [\DTLfetch{Moments (standard)}{Moment}{mean}{Lower_NMTA}; \DTLfetch{Moments (standard)}{Moment}{mean}{Upper_NMTA}] & \DTLfetch{Moments (adaptive)}{Moment}{mean}{GA} & [\DTLfetch{Moments (adaptive)}{Moment}{mean}{Lower_GA}; \DTLfetch{Moments (adaptive)}{Moment}{mean}{Upper_GA}] & \DTLfetch{Moments (adaptive)}{Moment}{mean}{NMTA} & [\DTLfetch{Moments (adaptive)}{Moment}{mean}{Lower_NMTA}; \DTLfetch{Moments (adaptive)}{Moment}{mean}{Upper_NMTA}] & \DTLfetch{Moments (adaptive)}{Moment}{mean}{Emperical} \\
        Standard deviation & \DTLfetch{Moments (standard)}{Moment}{stdev}{GA} & [\DTLfetch{Moments (standard)}{Moment}{stdev}{Lower_GA}; \DTLfetch{Moments (standard)}{Moment}{stdev}{Upper_GA}] & \DTLfetch{Moments (standard)}{Moment}{stdev}{NMTA} & [\DTLfetch{Moments (standard)}{Moment}{stdev}{Lower_NMTA}; \DTLfetch{Moments (standard)}{Moment}{stdev}{Upper_NMTA}] & \DTLfetch{Moments (adaptive)}{Moment}{stdev}{GA} & [\DTLfetch{Moments (adaptive)}{Moment}{stdev}{Lower_GA}; \DTLfetch{Moments (adaptive)}{Moment}{stdev}{Upper_GA}] & \DTLfetch{Moments (adaptive)}{Moment}{stdev}{NMTA} & [\DTLfetch{Moments (adaptive)}{Moment}{stdev}{Lower_NMTA}; \DTLfetch{Moments (adaptive)}{Moment}{stdev}{Upper_NMTA}] & \DTLfetch{Moments (adaptive)}{Moment}{stdev}{Emperical} \\
        Excess Kurtosis & \DTLfetch{Moments (standard)}{Moment}{kurtosis}{GA} & [\DTLfetch{Moments (standard)}{Moment}{kurtosis}{Lower_GA}; \DTLfetch{Moments (standard)}{Moment}{kurtosis}{Upper_GA}] & \DTLfetch{Moments (standard)}{Moment}{kurtosis}{NMTA} & [\DTLfetch{Moments (standard)}{Moment}{kurtosis}{Lower_NMTA}; \DTLfetch{Moments (standard)}{Moment}{kurtosis}{Upper_NMTA}] & \DTLfetch{Moments (adaptive)}{Moment}{kurtosis}{GA} & [\DTLfetch{Moments (adaptive)}{Moment}{kurtosis}{Lower_GA}; \DTLfetch{Moments (adaptive)}{Moment}{kurtosis}{Upper_GA}] & \DTLfetch{Moments (adaptive)}{Moment}{kurtosis}{NMTA} & [\DTLfetch{Moments (adaptive)}{Moment}{kurtosis}{Lower_NMTA}; \DTLfetch{Moments (adaptive)}{Moment}{kurtosis}{Upper_NMTA}] & \DTLfetch{Moments (adaptive)}{Moment}{kurtosis}{Emperical} \\
        Kolmogorov-Smirnov statistic & \DTLfetch{Moments (standard)}{Moment}{ks}{GA} & [\DTLfetch{Moments (standard)}{Moment}{ks}{Lower_GA}; \DTLfetch{Moments (standard)}{Moment}{ks}{Upper_GA}] & \DTLfetch{Moments (standard)}{Moment}{ks}{NMTA} & [\DTLfetch{Moments (standard)}{Moment}{ks}{Lower_NMTA}; \DTLfetch{Moments (standard)}{Moment}{ks}{Upper_NMTA}] & \DTLfetch{Moments (adaptive)}{Moment}{ks}{GA} & [\DTLfetch{Moments (adaptive)}{Moment}{ks}{Lower_GA}; \DTLfetch{Moments (adaptive)}{Moment}{ks}{Upper_GA}] & \DTLfetch{Moments (adaptive)}{Moment}{ks}{NMTA} & [\DTLfetch{Moments (adaptive)}{Moment}{ks}{Lower_NMTA}; \DTLfetch{Moments (adaptive)}{Moment}{ks}{Upper_NMTA}] & \DTLfetch{Moments (adaptive)}{Moment}{ks}{Emperical} \\
        Hurst exponent & \DTLfetch{Moments (standard)}{Moment}{hurst}{GA} & [\DTLfetch{Moments (standard)}{Moment}{hurst}{Lower_GA}; \DTLfetch{Moments (standard)}{Moment}{hurst}{Upper_GA}] & \DTLfetch{Moments (standard)}{Moment}{hurst}{NMTA} & [\DTLfetch{Moments (standard)}{Moment}{hurst}{Lower_NMTA}; \DTLfetch{Moments (standard)}{Moment}{hurst}{Upper_NMTA}] & \DTLfetch{Moments (adaptive)}{Moment}{hurst}{GA} & [\DTLfetch{Moments (adaptive)}{Moment}{hurst}{Lower_GA}; \DTLfetch{Moments (adaptive)}{Moment}{hurst}{Upper_GA}] & \DTLfetch{Moments (adaptive)}{Moment}{hurst}{NMTA} & [\DTLfetch{Moments (adaptive)}{Moment}{hurst}{Lower_NMTA}; \DTLfetch{Moments (adaptive)}{Moment}{hurst}{Upper_NMTA}] & \DTLfetch{Moments (adaptive)}{Moment}{hurst}{Emperical} \\
        GPH estimator & \DTLfetch{Moments (standard)}{Moment}{gph}{GA} & [\DTLfetch{Moments (standard)}{Moment}{gph}{Lower_GA}; \DTLfetch{Moments (standard)}{Moment}{gph}{Upper_GA}] & \DTLfetch{Moments (standard)}{Moment}{gph}{NMTA} & [\DTLfetch{Moments (standard)}{Moment}{gph}{Lower_NMTA}; \DTLfetch{Moments (standard)}{Moment}{gph}{Upper_NMTA}] & \DTLfetch{Moments (adaptive)}{Moment}{gph}{GA} & [\DTLfetch{Moments (adaptive)}{Moment}{gph}{Lower_GA}; \DTLfetch{Moments (adaptive)}{Moment}{gph}{Upper_GA}] & \DTLfetch{Moments (adaptive)}{Moment}{gph}{NMTA} & [\DTLfetch{Moments (adaptive)}{Moment}{gph}{Lower_NMTA}; \DTLfetch{Moments (adaptive)}{Moment}{gph}{Upper_NMTA}] & \DTLfetch{Moments (adaptive)}{Moment}{gph}{Emperical} \\
        ADF statistic & \DTLfetch{Moments (standard)}{Moment}{adf}{GA} & [\DTLfetch{Moments (standard)}{Moment}{adf}{Lower_GA}; \DTLfetch{Moments (standard)}{Moment}{adf}{Upper_GA}] & \DTLfetch{Moments (standard)}{Moment}{adf}{NMTA} & [\DTLfetch{Moments (standard)}{Moment}{adf}{Lower_NMTA}; \DTLfetch{Moments (standard)}{Moment}{adf}{Upper_NMTA}] & \DTLfetch{Moments (adaptive)}{Moment}{adf}{GA} & [\DTLfetch{Moments (adaptive)}{Moment}{adf}{Lower_GA}; \DTLfetch{Moments (adaptive)}{Moment}{adf}{Upper_GA}] & \DTLfetch{Moments (adaptive)}{Moment}{adf}{NMTA} & [\DTLfetch{Moments (adaptive)}{Moment}{adf}{Lower_NMTA}; \DTLfetch{Moments (adaptive)}{Moment}{adf}{Upper_NMTA}] & \DTLfetch{Moments (adaptive)}{Moment}{adf}{Emperical} \\
        GARCH parameters & \DTLfetch{Moments (standard)}{Moment}{garch}{GA} & [\DTLfetch{Moments (standard)}{Moment}{garch}{Lower_GA}; \DTLfetch{Moments (standard)}{Moment}{garch}{Upper_GA}] & \DTLfetch{Moments (standard)}{Moment}{garch}{NMTA} & [\DTLfetch{Moments (standard)}{Moment}{garch}{Lower_NMTA}; \DTLfetch{Moments (standard)}{Moment}{garch}{Upper_NMTA}] & \DTLfetch{Moments (adaptive)}{Moment}{garch}{GA} & [\DTLfetch{Moments (adaptive)}{Moment}{garch}{Lower_GA}; \DTLfetch{Moments (adaptive)}{Moment}{garch}{Upper_GA}] & \DTLfetch{Moments (adaptive)}{Moment}{garch}{NMTA} & [\DTLfetch{Moments (adaptive)}{Moment}{garch}{Lower_NMTA}; \DTLfetch{Moments (adaptive)}{Moment}{garch}{Upper_NMTA}] & \DTLfetch{Moments (adaptive)}{Moment}{garch}{Emperical} \\
        Hill estimator & \DTLfetch{Moments (standard)}{Moment}{hill}{GA} & [\DTLfetch{Moments (standard)}{Moment}{hill}{Lower_GA}; \DTLfetch{Moments (standard)}{Moment}{hill}{Upper_GA}] & \DTLfetch{Moments (standard)}{Moment}{hill}{NMTA} & [\DTLfetch{Moments (standard)}{Moment}{hill}{Lower_NMTA}; \DTLfetch{Moments (standard)}{Moment}{hill}{Upper_NMTA}] & \DTLfetch{Moments (adaptive)}{Moment}{hill}{GA} & [\DTLfetch{Moments (adaptive)}{Moment}{hill}{Lower_GA}; \DTLfetch{Moments (adaptive)}{Moment}{hill}{Upper_GA}] & \DTLfetch{Moments (adaptive)}{Moment}{hill}{NMTA} & [\DTLfetch{Moments (adaptive)}{Moment}{hill}{Lower_NMTA}; \DTLfetch{Moments (adaptive)}{Moment}{hill}{Upper_NMTA}] & \DTLfetch{Moments (adaptive)}{Moment}{hill}{Emperical} \\ \bottomrule
    \end{tabular}
    \caption{\label{table:Moments and Statistics} Moments and statistics on actual vs simulated data using the GA and NMTA optimisation methods}
\end{sidewaystable}

\subsection{Simulation}
\begin{figure*}
\begin{minipage}{.48\textwidth}
    \begin{figure}[H]
        \centering
        \begin{subfigure}[t]{\textwidth}
            \centering
            \includegraphics[trim = 0 6 105 10, clip, width = \textwidth]{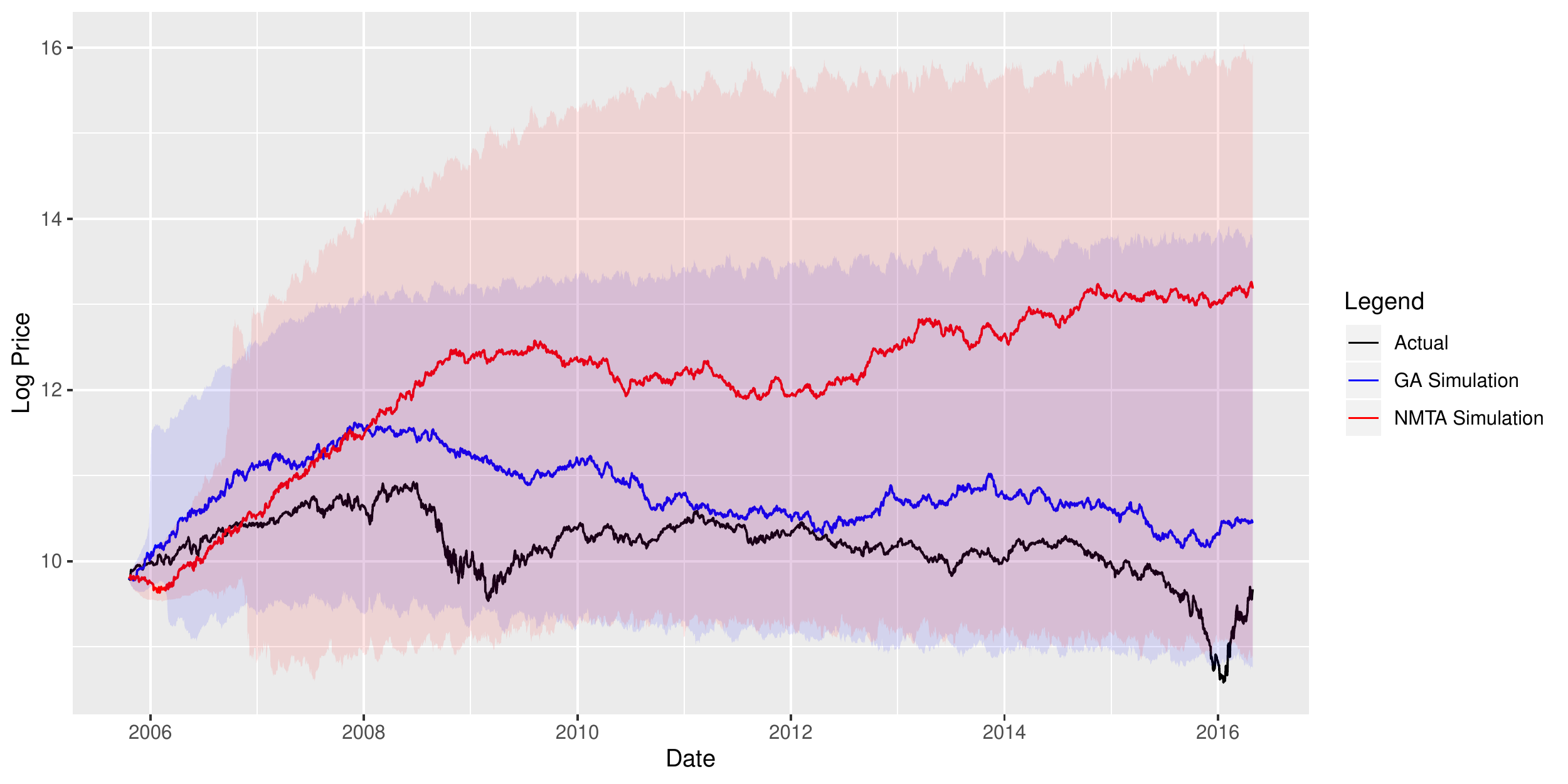}
            \caption{Closing log-price paths \label{fig:SFJ - Price paths}}
        \end{subfigure}
        \begin{subfigure}[t]{0.49\textwidth}
            \includegraphics[width = \textwidth]{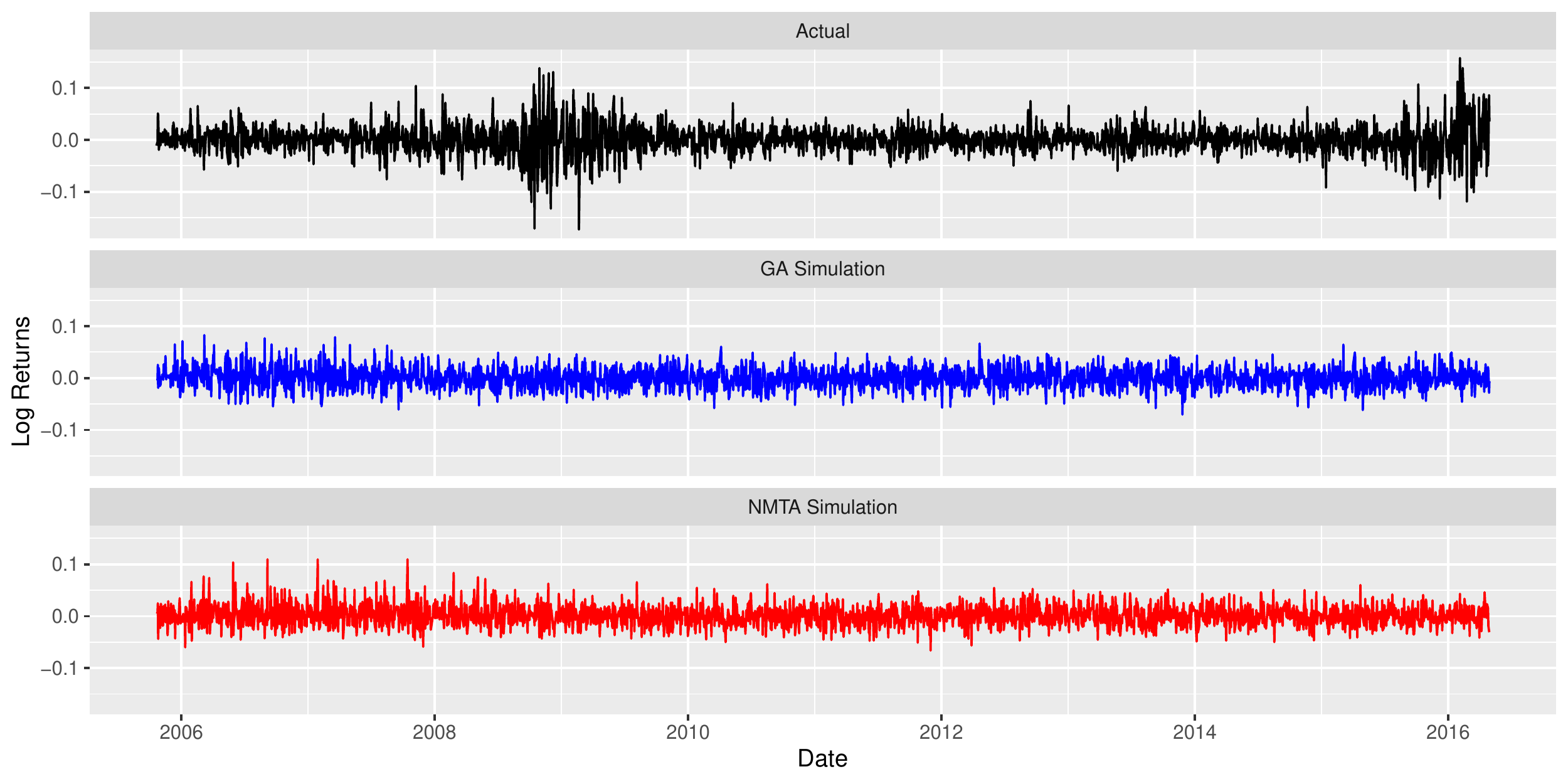}
            \caption{Log-return paths \label{fig:SFJ - Return paths}}
        \end{subfigure}
        \begin{subfigure}[t]{0.49\textwidth}
            \includegraphics[width = \textwidth]{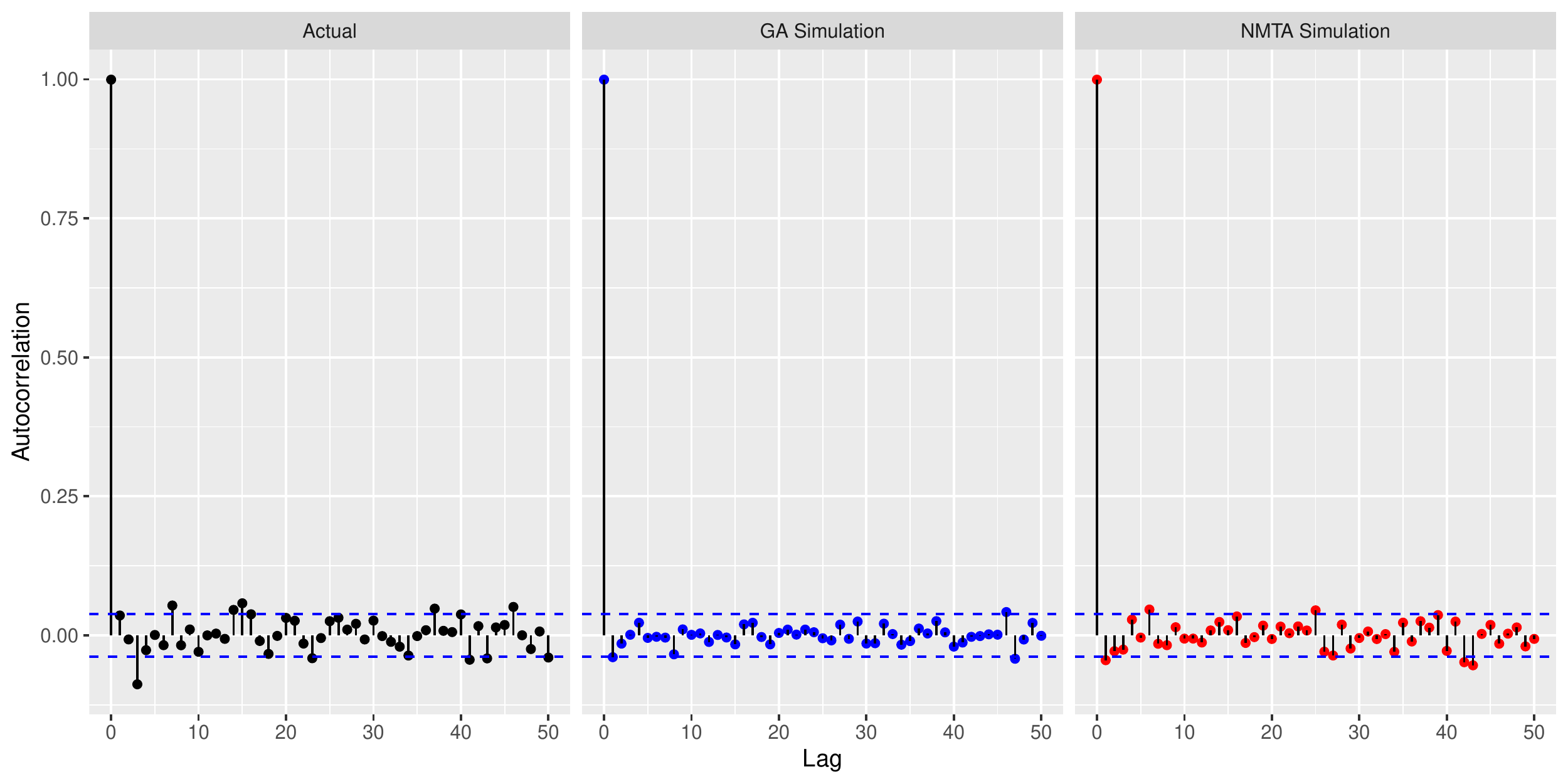}
            \caption{Auto-correlation of $r_t$ \label{fig:SFJ - Autocorrelation of log returns}}
        \end{subfigure}
        \begin{subfigure}[t]{0.49\textwidth}
            \includegraphics[width = \textwidth]{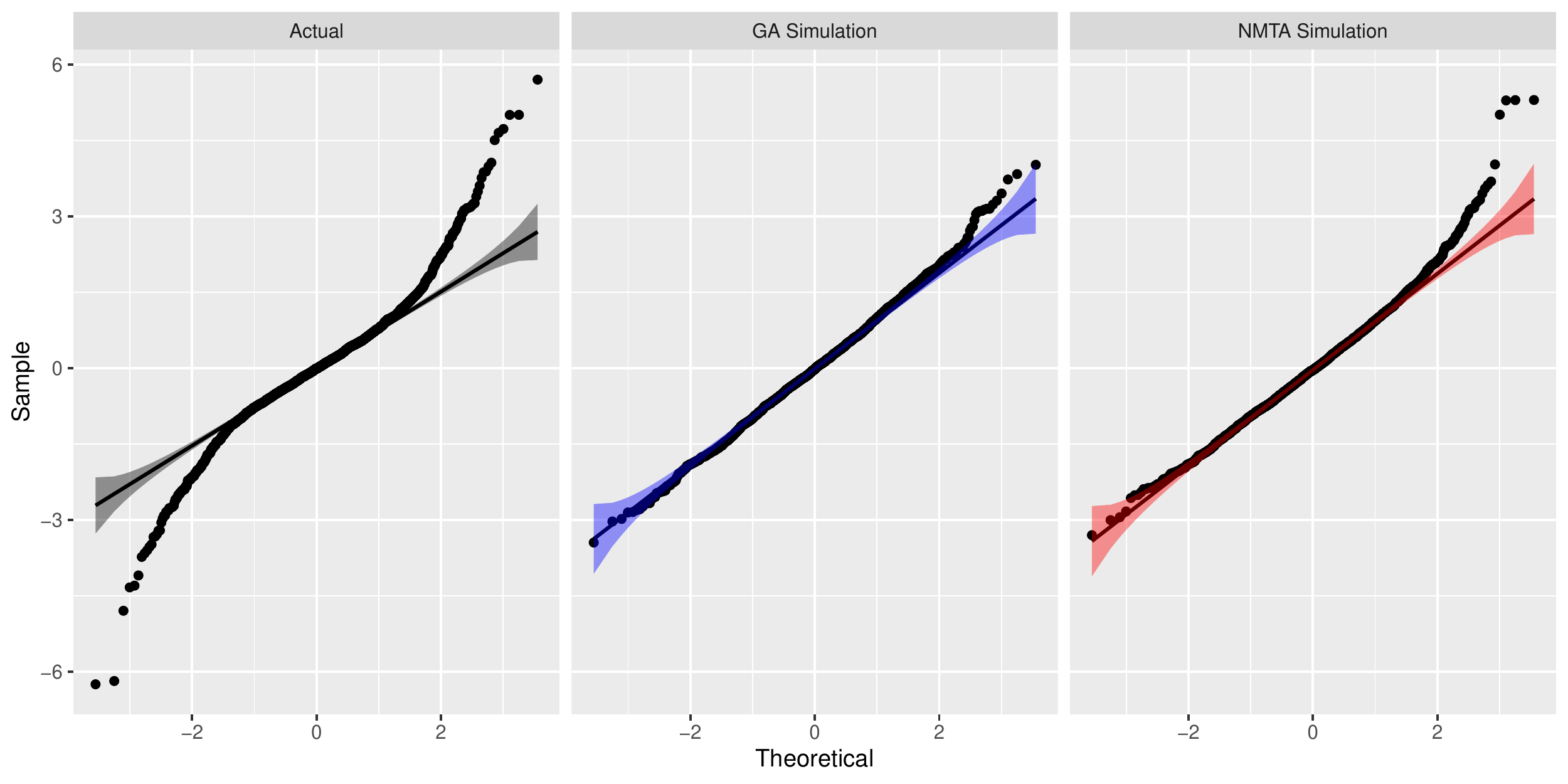}
            \caption{QQ plots of $r_t$ relative to the normal distribution \label{fig:SFJ - Normal probability plots}}
        \end{subfigure}
        \begin{subfigure}[t]{0.49\textwidth}
            \includegraphics[width = \textwidth]{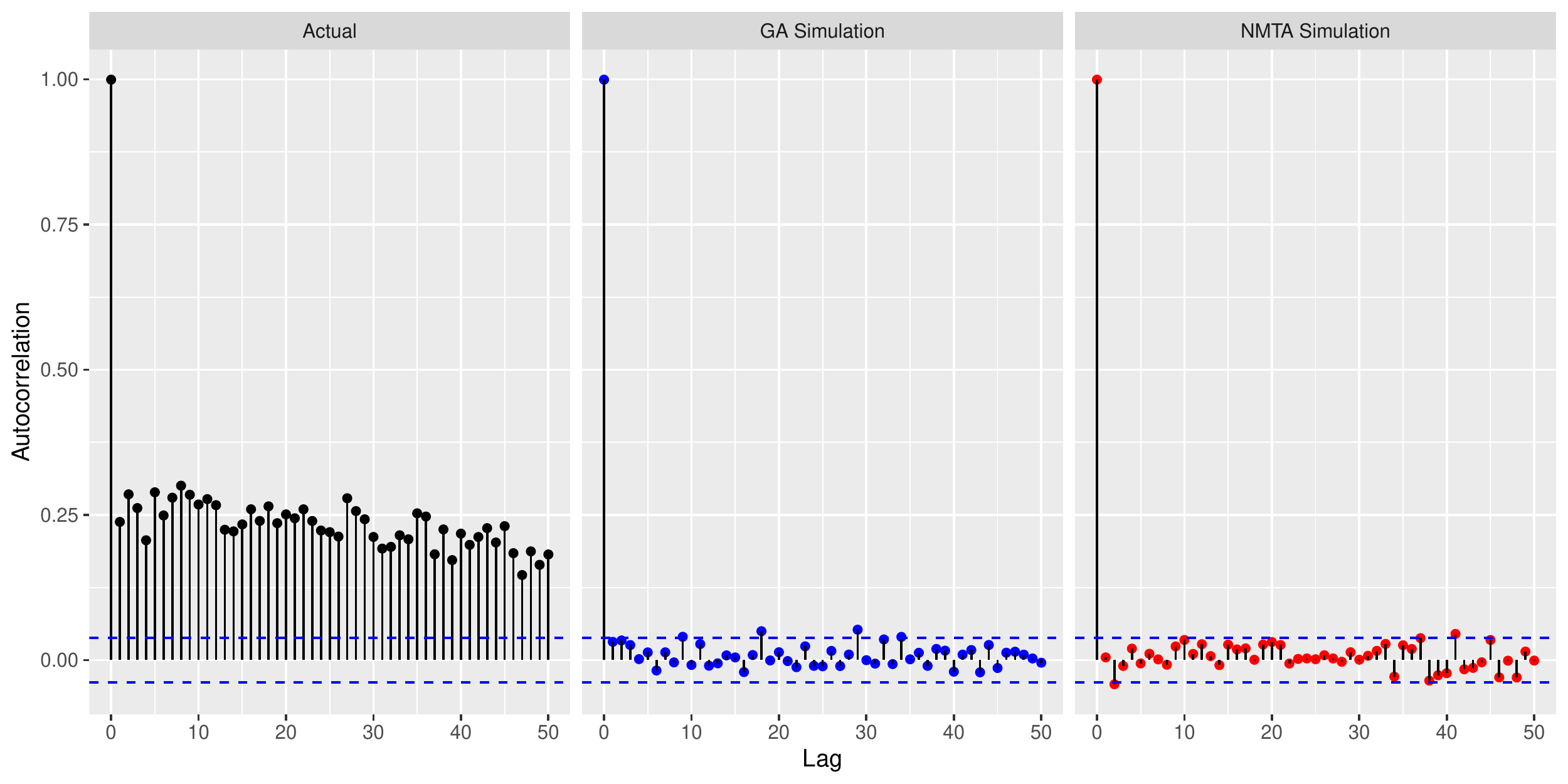}
            \caption{Auto-correlation of $|r_t|$ \label{fig:SFJ - Autocorrelation of absolute log returns}}
        \end{subfigure}
        \caption{Actual features (black) vs features generated from simulations of the standard model using best GA (blue) and NMTA (red) results \label{fig:SFJ-Results}}
    \end{figure}
\end{minipage}
\begin{minipage}{.48\textwidth}
    \begin{figure}[H]
        \centering
        \begin{subfigure}[t]{\textwidth}
            \centering
            \includegraphics[trim = 0 6 105 10, clip, width = \textwidth]{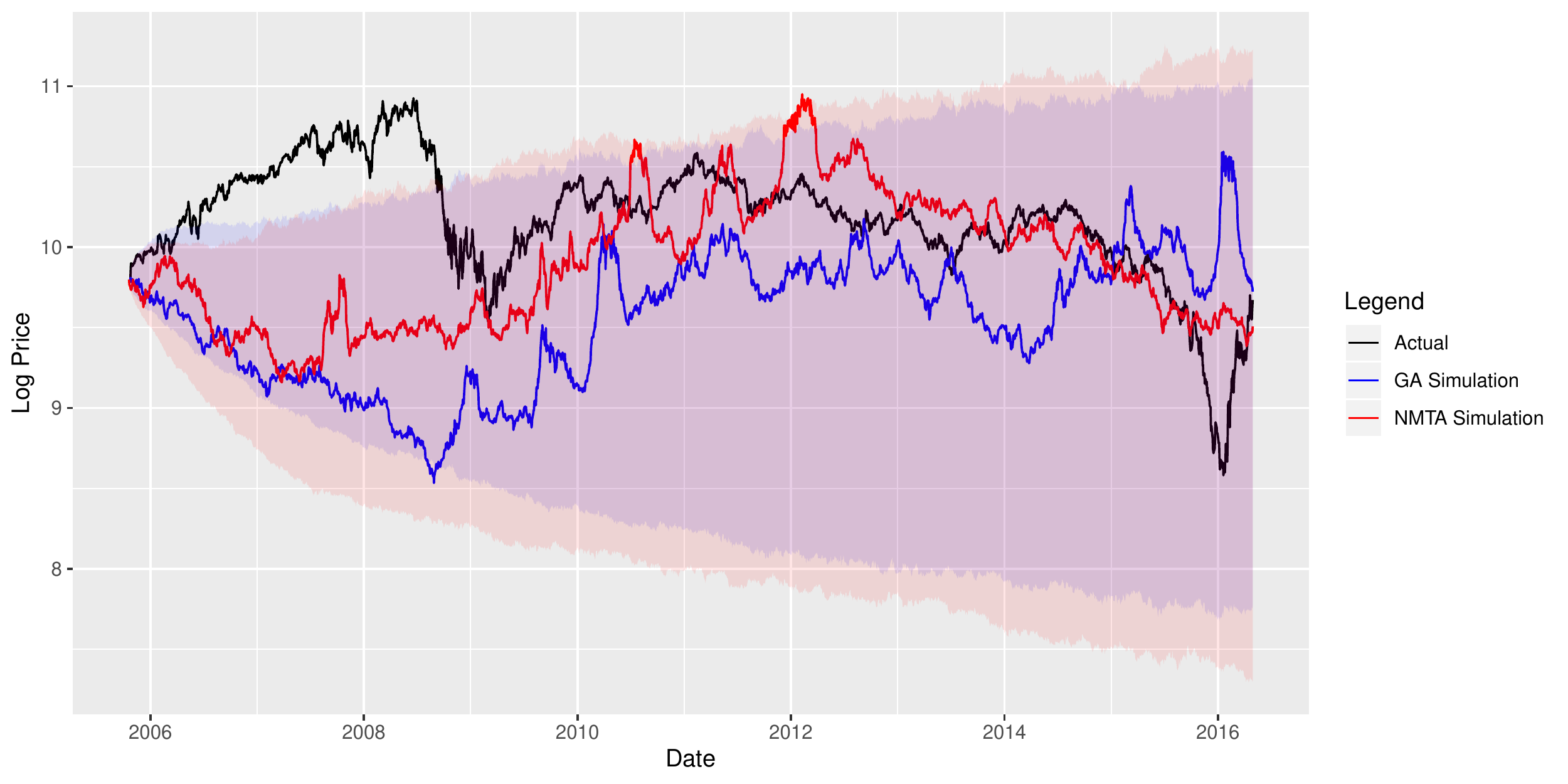}
            \caption{Closing log-price paths \label{fig:AFJ - Price paths}}
        \end{subfigure}
        \begin{subfigure}[t]{0.49\textwidth}
            \includegraphics[width = \textwidth]{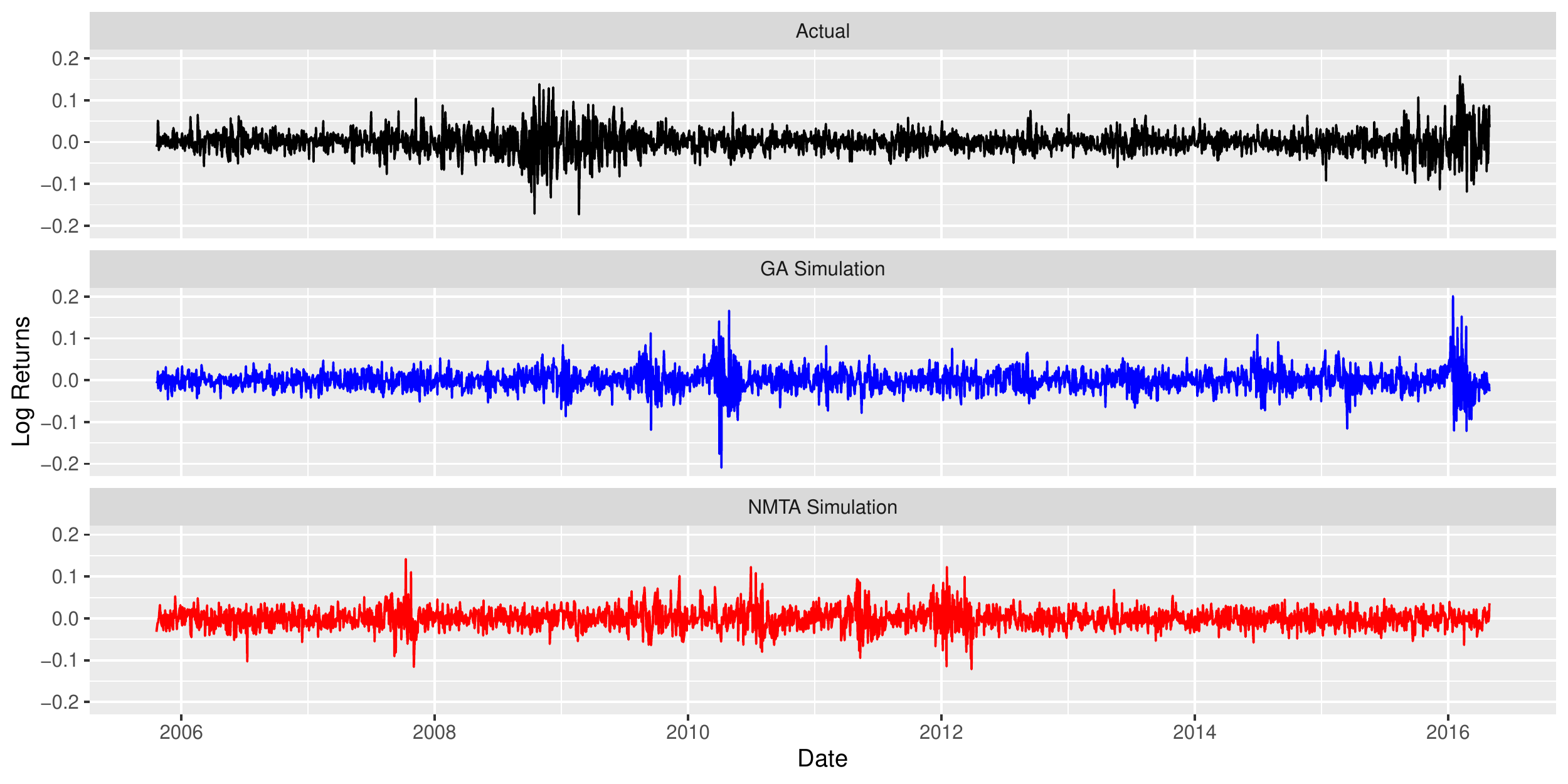}
            \caption{Log-return paths \label{fig:AFJ - Return paths}}
        \end{subfigure}
        \begin{subfigure}[t]{0.49\textwidth}
            \includegraphics[width = \textwidth]{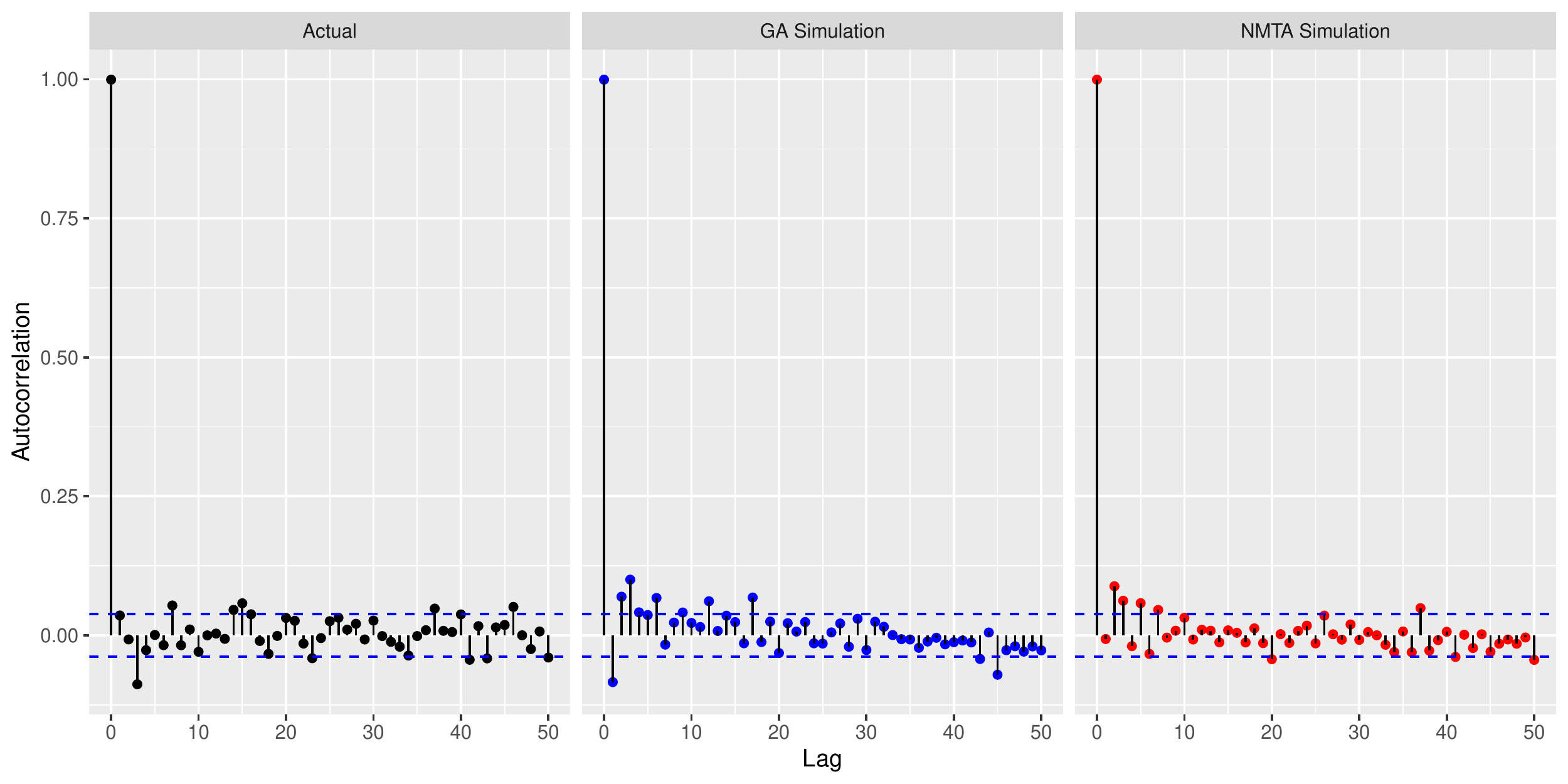}
            \caption{Auto-correlation of $r_t$ \label{fig:AFJ - Autocorrelation of log returns}}
        \end{subfigure}
        \begin{subfigure}[t]{0.49\textwidth}
            \includegraphics[width = \textwidth]{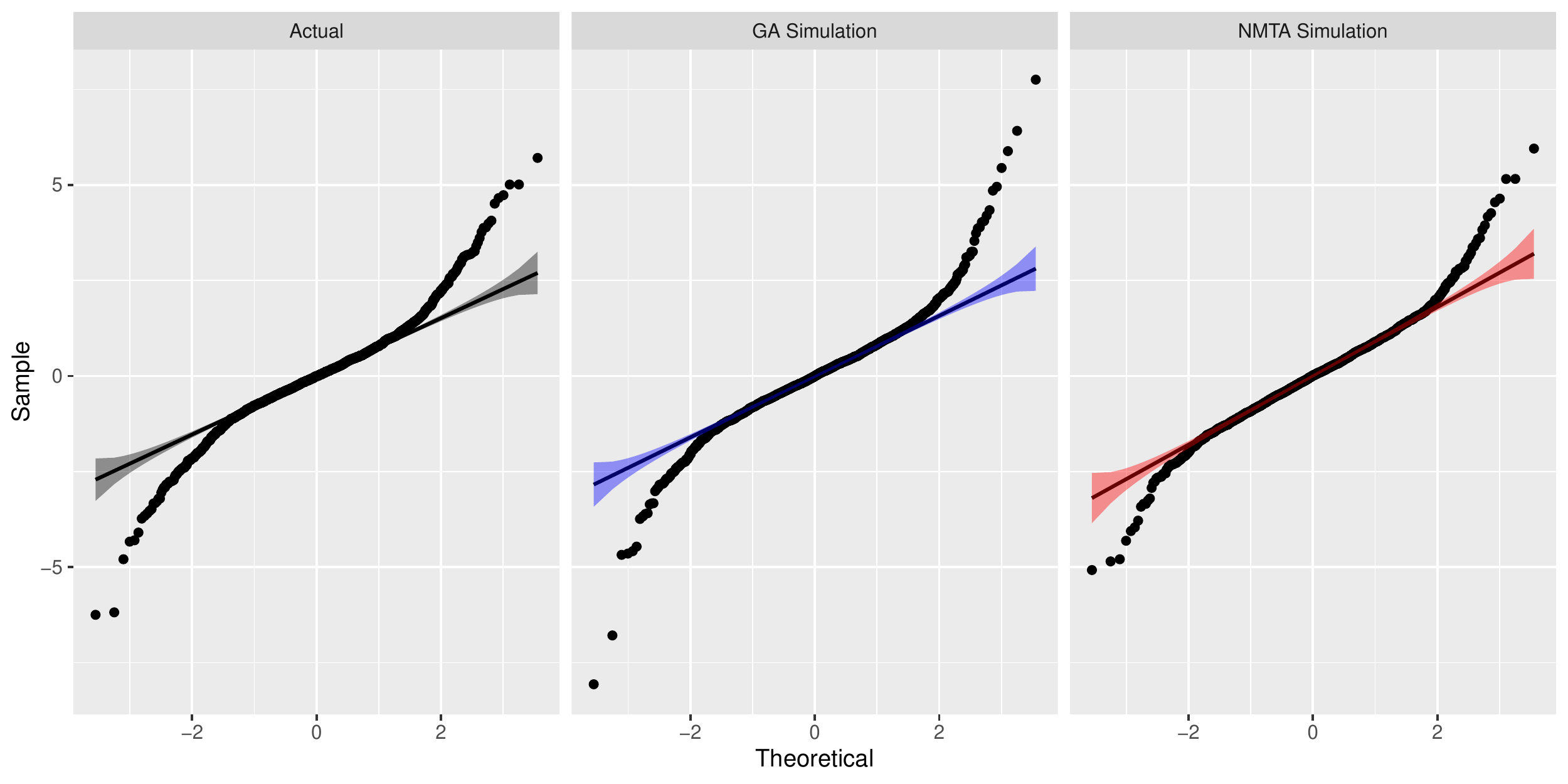}
            \caption{QQ plots of $r_t$ relative to the normal distribution \label{fig:AFJ - Normal probability plots}}
        \end{subfigure}
        \begin{subfigure}[t]{0.49\textwidth}
            \includegraphics[width = \textwidth]{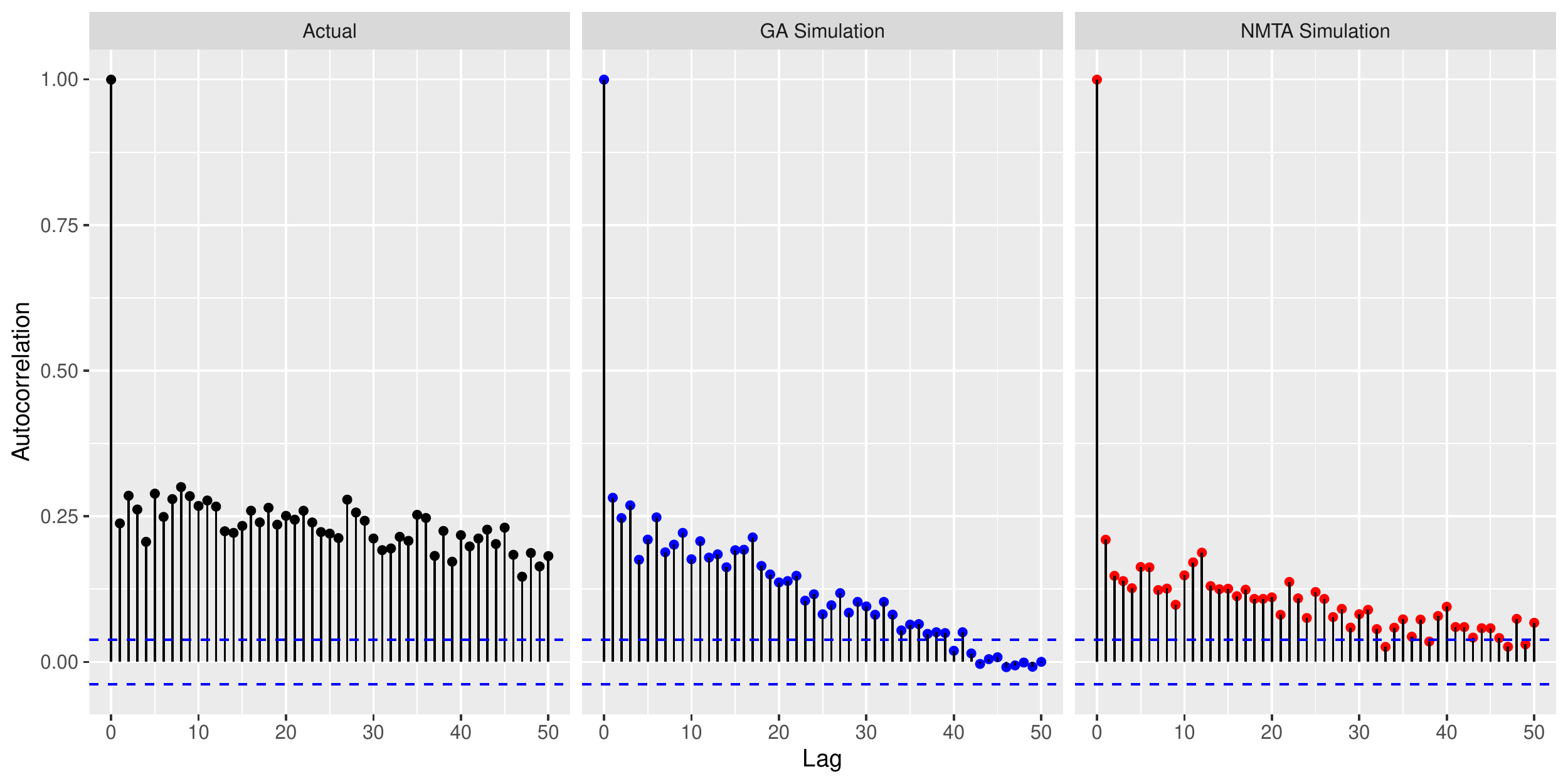}
            \caption{Auto-correlation of $|r_t|$ \label{fig:AFJ - Autocorrelation of absolute log returns}}
        \end{subfigure}
        \caption{Actual features (black) vs features generated from simulations of the adaptive model using best GA (blue) and NMTA (red) results \label{fig:AFJ-Results}}
    \end{figure}
\end{minipage}
\end{figure*}

Figures \ref{fig:SFJ-Results} and \ref{fig:AFJ-Results} compare the observed data to the simulations using the best-performing parameter values from the GA and NMTA calibrations. Figure \ref{fig:SFJ - Return paths} shows that the variance of returns for the standard model simulations are very consistent compared to the observed data. We find minimal fat tails (figure \ref{fig:SFJ - Normal probability plots}) in the simulations. The simulations replicate the lack of autocorrelations of log returns found in the data, but they fail to produce the observed autocorrelations of absolute log returns (figures \ref{fig:SFJ - Autocorrelation of log returns} and \ref{fig:SFJ - Autocorrelation of absolute log returns}). Overall, as with a previous calibration attempt by \citet{fabretti2013problem}, we could not replicate some important stylized facts well.

For the adaptive model, we observe greater consistency between simulations, shown by narrower price confidence intervals (shaded regions). The daily log returns more closely resemble actual log returns, but clustered volatility is not as prominent as the actual data (figure \ref{fig:AFJ - Return paths}). Figure \ref{fig:AFJ - Normal probability plots} shows that both calibrations generate fat tails, with those from the NMTA results being less prominent. Both have almost no significant autocorrelations of log returns. The GA calibration produces highly significant autocorrelations of absolute log returns that decay at a faster rate than the observed data, whilst the NMTA's autocorrelations are less pronounced but decay more slowly. Overall, the adaptive model much more closely mimics the real data.

Observing the number of active chartists and fundamentalists for the adaptive model at each time step reveals that switching is relatively volatile and correspond to areas of volatility clustering in the plot of log returns. For this reason, we conclude that a large fraction of fundamentalists tends to stabilise prices, whereas a large fraction of chartists tends to destabilise prices. Furthermore, asset price fluctuations are caused by the interaction between these stabilizing and destabilizing forces.

\begin{figure}[H]
    \centering
    \includegraphics[width=\linewidth]{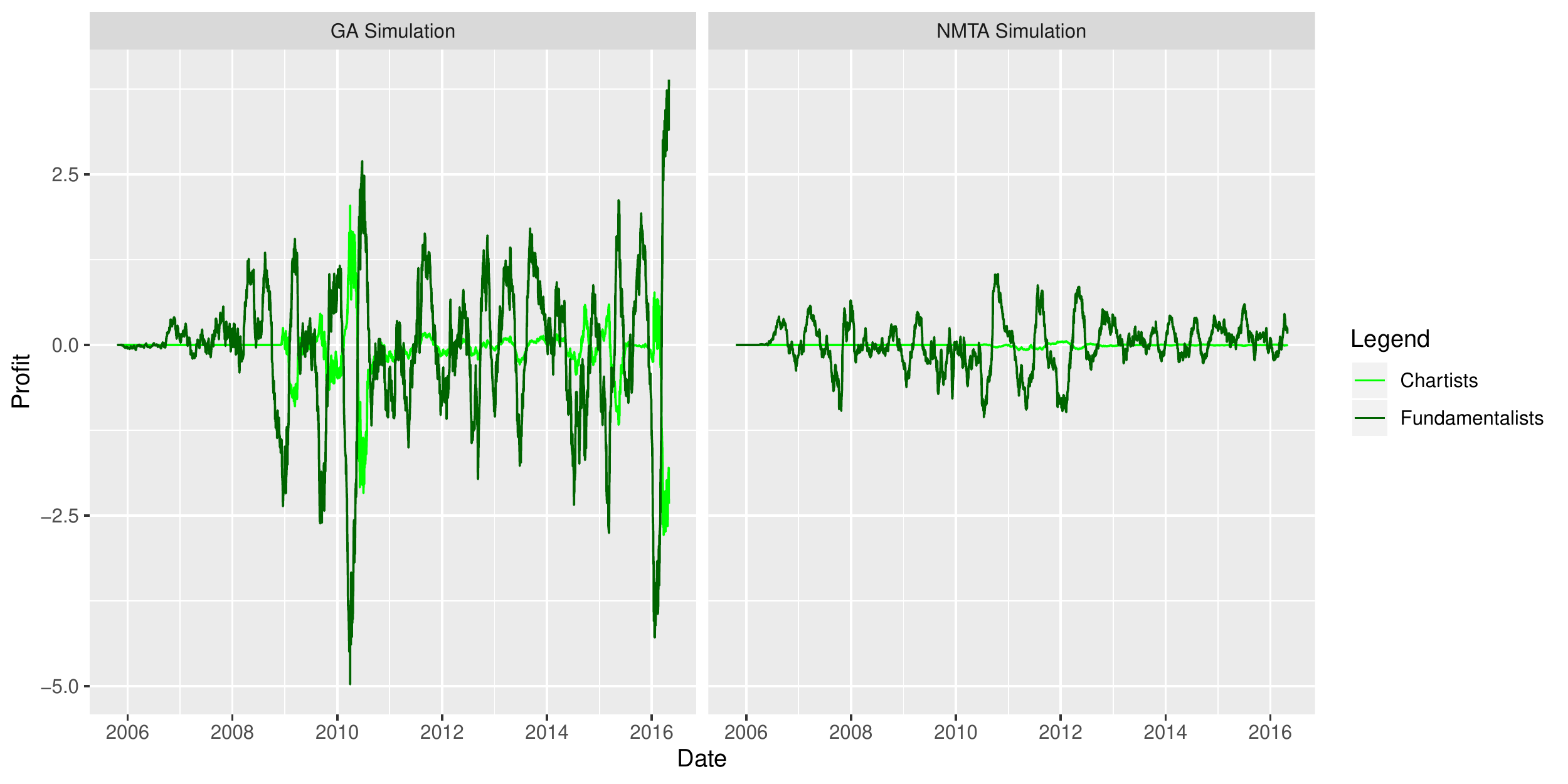}
    \caption{\label{fig:Profit of chartists/fundamentalists}Profit of chartists/fundamentalists}
\end{figure}

\section{Conclusion and Criticisms}
The less leptokurtic results from the standard model is likely not due to the optimisation procedures but to the model characteristics, as the same procedures achieved better results from the adaptive model. A key feature of the adaptive model is that the fluctuations in the number of chartists and fundamentalists are triggered by a rational choice, based upon maximising profits over a time horizon of two to three months. As in \citet{brock1998heterogeneous}, it was found that the ``intensity of choice'' to switch predictors can lead to market instability and the emergence of complicated dynamics for asset prices and returns. When the intensity of switching is high, chartist activity fluctuates violently and asset price fluctuations are indeed characterized by an irregular switching between phases where prices are close to the EMH fundamental price behaviour, phases of optimism, and phases of pessimism \cite{brock1998heterogeneous}. We also recover the well understood result that a large fraction of fundamentalists tends to stabilise prices, whilst a large fraction of chartists tends to destabilise prices. However, we observe relatively large confidence intervals and haphazard or flat objective function surfaces for many parameters. This indicates that they do not have a very clear effect on price behaviour, making it difficult to draw insights -- both models exhibit parameter degeneracies - that is, independent calibrations on the same data do not yield similar optimal parameters, but do have comparable ability to replicate the stylized facts. So the problem remains that broadly the dynamics are interesting and perhaps reasonable, but linking these robustly to particular parameter values is problematic. We do not claim realism in our results. However, with a simply feedback in terms of monitoring strategy profitability and then making strategy switching decisions based on past profitability, as monitored on an at least a monthly horizon, we can reproduce auto-correlations in the absolute values of the price fluctuations along with enhanced tail event activity.   

\section{Acknowledgements}
We would like to thank members of the Statistical Finance Research Group at UCT - in particular, Patrick Chang, Melusi Mavuso, Lionel Yelibi, and  Etienne Pienaar - for their helpful feedback, probing questions, and constructive criticism.

\bibliographystyle{elsarticle-num-names}
\bibliography{AdaptiveFarmerJoshi}

\onecolumn
\appendix

\end{document}

%% file: Preamble.tex
\usepackage{float} 
\usepackage{comment} 
\usepackage[space]{grffile} 
\graphicspath{{Figures/}}
\usepackage{subcaption} 
\usepackage[colorlinks=true, allcolors=black]{hyperref} 
\usepackage{siunitx} 
\usepackage{wrapfig}
\usepackage{bm}
\usepackage{rotating}
\usepackage{booktabs}
\usepackage{datatool} 
\usepackage{graphicx}
\DTLsetseparator{,}
\DTLloaddb{Moments (adaptive)}{"Data/AFJ-Moments.txt"}
\DTLloaddb{Parameters (adaptive)}{"Data/AFJ-Parameters.txt"}
\DTLloaddb{Moments (standard)}{"Data/SFJ-Moments.txt"}
\DTLloaddb{Parameters (standard)}{"Data/SFJ-Parameters.txt"}
\makeatletter
\def\ps@pprintTitle{%
    \let\@oddhead\@empty
    \let\@evenhead\@empty
    \let\@oddfoot\@empty
    \let\@evenfoot\@oddfoot
}
\makeatother